\documentclass[letterpaper]{article}

\usepackage[T1]{fontenc}

\usepackage{geometry}
\usepackage{setspace}

\usepackage[style = chem-acs,articletitle=true,autocite=superscript]{biblatex}
\usepackage{graphicx}
\usepackage{float}
\newfloat{scheme}{htbp}{los}
\floatname{scheme}{Scheme}
\floatname{chart}{Chart}
\newfloat{graph}{htbp}{loh}

\usepackage{chemformula} 
\usepackage[version = 4]{mhchem} 

\usepackage{amsmath,amssymb}

\usepackage{soul}

\usepackage{authblk}
\author[1,2,3]{Aurelian Loirette-Pelous*}
\author[2,4,5]{Roberto A. Boto}
\author[2,3,5]{Javier Aizpurua}
\author[1,2]{Ruben Esteban*}
\affil[1]{Centro de Física de Materiales (CFM-MPC), CSIC-EHU), Paseo Manuel de Lardizabal 5, 20018 Donostia-San Sebastián, Spain}%
\affil[2]{Donostia International Physics Center (DIPC), Paseo Manuel de Lardizabal 4, 20018 Donostia-San Sebastián, Spain}%
\affil[3]{Department of Electricity and Electronics, FCT-ZTF, UPV-EHU, Bilbao, 48080, Spain}%
\affil[4]{Department of Polymers and Advanced Materials, Faculty of Chemistry, The University of the Basque Country, UPV/EHU, 20018 Donostia-San Sebastián, Euskadi, Spain}%
\affil[5]{IKERBASQUE, Basque Foundation for Science, 48009 Bilbao, Spain}%

\title{Addressing intramolecular vibrational redistribution in a single molecule through pump and probe surface-enhanced vibrational spectroscopy}
\date{*E-mail: aurelianjoan.loirettepelous@ehu.eus; ruben.esteban@ehu.eus}

\begin{document}

\maketitle

\begin{abstract}
The development of accurate tools to characterize Intramolecular Vibrational Redistribution (IVR) is of major interest in chemistry. In this context, surface-enhanced vibrational spectroscopies stand up as well-established techniques to study molecular vibrational lines and populations with a sensitivity that can reach the single-molecule level. However, to date, this possibility has not been fully developed to address IVR. Here, we establish a quantum mechanical framework based on molecular optomechanics that accounts for IVR, and we adopt it to analyze strategies to optimize IVR characterization by surface-enhanced vibrational spectroscopy. In particular, we model two different pump-and-probe configurations where the vibrational pumping is provided either by infrared laser illumination or by Stokes scattering processes in surface-enhanced Raman spectroscopy (SERS). We show for the two pumping configurations the existence of clear signatures on the anti-Stokes SERS spectra of population transfer between coupled vibrational modes in a molecule. Our calculations adopt realistic molecular and SERS parameters, suggesting that these signatures of IVR could be accessible at the single-molecule level within realistic experimental platforms.
\end{abstract}

\section*{Keywords}

Pump-and-Probe Vibrational Spectroscopy, Surface-Enhanced Raman Scattering, Intramolecular Vibrational Redistribution, Molecular Optomechanics



\section{Introduction}\label{sec:introduction} 

A long-sought goal in chemistry is to control chemical reactions by exciting selectively a few vibrational modes of a reactant molecule \autocite{zewail1980laser,warren1993coherent,zare1998laser}. However, this concept has proven challenging to implement due to the rapid redistribution of locally deposited energy throughout the molecule. In particular, Intramolecular Vibrational Redistribution (IVR), which refers to population redistribution within a given vibrational manifold, is the fastest relaxation stage with a typical 0.1-10 ps timescale \autocite{freed1980intramolecular,nesbitt1996vibrational,gruebele2004vibrational}. As sketched in Fig. \ref{fig:schema} (bottom inset), IVR is mainly due to the resonant coupling of different combinations of vibrational modes, where the coupling originates from anharmonicities of the molecular potential energy surface \autocite{may2023charge}. Interestingly, recent years have seen the advent of approaches to manipulate molecular potential energy surfaces, as e.g., strong-coupling between vibrations and vacuum electromagnetic fields \autocite{thomas2019tilting,nagarajan2021chemistry,ahn2023modification,xiang2024molecular} or the formation of charge transfer states between a molecule and a metal in plasmonic nanostructures \autocite{zhang2017surface,zhan2018plasmon,zhan2020recent}. These approaches open new avenues for chemistry and further emphasize the interest of a detailed understanding of IVR to achieve their potential. However, experimental as well as theoretical identification of IVR pathways remains a notoriously difficult task in molecules with more than a handful of atoms. Therefore, the development of new IVR characterization tools is key to achieving efficient control of IVR and, ultimately, the success of mode-selective chemistry.

Optical time-resolved pump-and-probe spectroscopy are well established techniques to study IVR \autocite{fayer2013ultrafast}. These techniques enable the study of IVR both in the ground \autocite{rubtsov2019ballistic} and excited \autocite{brinks2014ultrafast} electronic states of molecules. In excited electronic states, IVR can be studied by pumping with visible or ultraviolet light and detecting the molecular fluorescence. The sensitivity of these pump-probe techniques is high, such that measurements on single molecules have been achieved \autocite{brinks2014ultrafast}. In this work, instead, we address the sensitivity of IVR characterization in the ground state. The common starting point of the existing pump-probe techniques of IVR in the ground state consists in pumping a given vibration away from thermal equilibrium. This is usually achieved using ultrafast pico- to femtosecond infrared (IR) laser pulses. Vibrational redistribution then induces a sizable population excess in a different vibrational mode that can be probed with light. In particular, a simple and broadband probing technique was historically provided by anti-Stokes Raman spectroscopy \autocite{laubereau1978vibrational,dlott2001vibrational}, which can reach sub-picosecond time resolution \autocite{deak1998vibrational,wang2002watching}. Better time and spectral resolutions were later achieved using more complex coherent techniques such as ultrafast two-dimensional infrared spectroscopy \autocite{khalil2003coherent,kuhs2019recent,farmer2024spectroscopy}. However, to date, these techniques are hindered by a low sensitivity, restricting studies to large ensembles of molecules \autocite{rubtsov2019ballistic}. Given that large molecular ensembles are prone to prominent intermolecular or molecule-solvent interactions \autocite{owrutsky1994vibrational} as well as to inhomogeneous effects, the accuracy of unambiguous IVR characterization is thus limited.

Optical nanoresonators provide a natural approach to achieve better sensitivity in vibrational spectroscopy, as they can generate large electric field enhancements and confinement that boost light-molecule interaction. This is the basic principle underpinning surface-enhanced infrared absorption \autocite{kozuch2023surface} and Raman \autocite{langer2019present} spectroscopies. In particular, surface-enhanced Raman spectroscopy (SERS) relying on plasmonic resonances of metallic nanoresonators is a well established technique for molecular detection down to the single molecule level \autocite{kneipp1997single,nie1997probing,xu1999spectroscopy,le2012single}. In the past decade, time-resolved pump-and-probe methods applied to SERS such as surface-enhanced coherent anti-Stokes Raman scattering \autocite{yampolsky2014seeing,jakob2024accelerated} or surface-enhanced femtosecond stimulated Raman spectroscopy \autocite{frontiera2011surface,buchanan2016surface} have also been able to approach or even reach the single-molecule limit. However, to date, the possibility to characterize IVR using surface-enhanced vibrational spectroscopy has been little explored, both experimentally and theoretically. In particular, it still remains unclear, to the best of our knowledge, whether characterization of IVR in a single molecule could be accessed within existing experimental platforms.

In this article, we establish a theoretical framework to model IVR and its characterization through surface-enhanced vibrational spectroscopy, which is based on a molecular optomechanical approach that captures the Raman interaction between nanoresonator photons and molecular vibrations \autocite{roelli2016molecular,schmidt2016quantum,esteban2022molecular,roelli2024nanocavities}. We then use this framework to explore promising experimental detection configurations, and provide evidence that realistic  SERS platforms based on ultrathin gap plasmonic nanocavities could enable the characterization of individual microscopic IVR pathways down to the single molecule limit. Specifically, we consider Fermi resonance couplings (one of the most common and efficient IVR pathway \autocite{may2023charge,poudel2024vibrational}) in a single molecule, and characterize them using a pump-and-probe scheme where probing is implemented by anti-Stokes non-resonant SERS. As sketched in Fig. \ref{fig:schema}, we specifically explore two pumping configurations: (i) Pumping with visible laser light of an optimized SERS platform, which can result in a large out-of-equilibrium driving of a vibration induced by successive Stokes Raman processes (vibrational pumping \autocite{galloway2009single}). Under pulsed excitation, we identify clear signatures of IVR that may be detected in a single or a few molecules. (ii) Pumping with infrared laser light a cavity that is resonantly coupled to the vibration; the IR modes of this cavity complement those at visible energies enhancing the Raman signal. In contrast to previous studies of IVR characterization based on pulsed IR pumping \autocite{laubereau1972direct,deak1998vibrational,wang2002watching}, the presence of the infrared cavity enables reaching large vibrational populations at low incident powers, showing that signatures of IVR of single molecules could also be observed under continuous-wave IR pumping. Finally, we discuss the assumptions made in our work to highlight that our approach is general and could be implemented within  realistic experimental platforms.

The article is organized as follows. In section \ref{sec:model}, we introduce the theoretical model to address IVR characterization within SERS. In section \ref{sec:results}, we analyze the different configurations to study the signatures of population transfer by IVR between two coupled vibrations. Realistic parameters to model these configurations are introduced in Section \ref{sec:parameters}. We then consider continuous-wave (Section \ref{sec:CW_IVR_SERS}) and pulsed (Section \ref{sec:pulsed_SERS}) vibrational pumping induced by Stokes SERS processes. Pumping with continuous-wave infrared light is studied in Section \ref{sec:IR_Raman_CW}. We finally summarize our results and discuss the generality of the approach in Section \ref{sec:discussion}.

\section{Theoretical framework}\label{sec:model}

\begin{figure}[t]
    \centering
    \includegraphics[width= 0.35\textwidth]{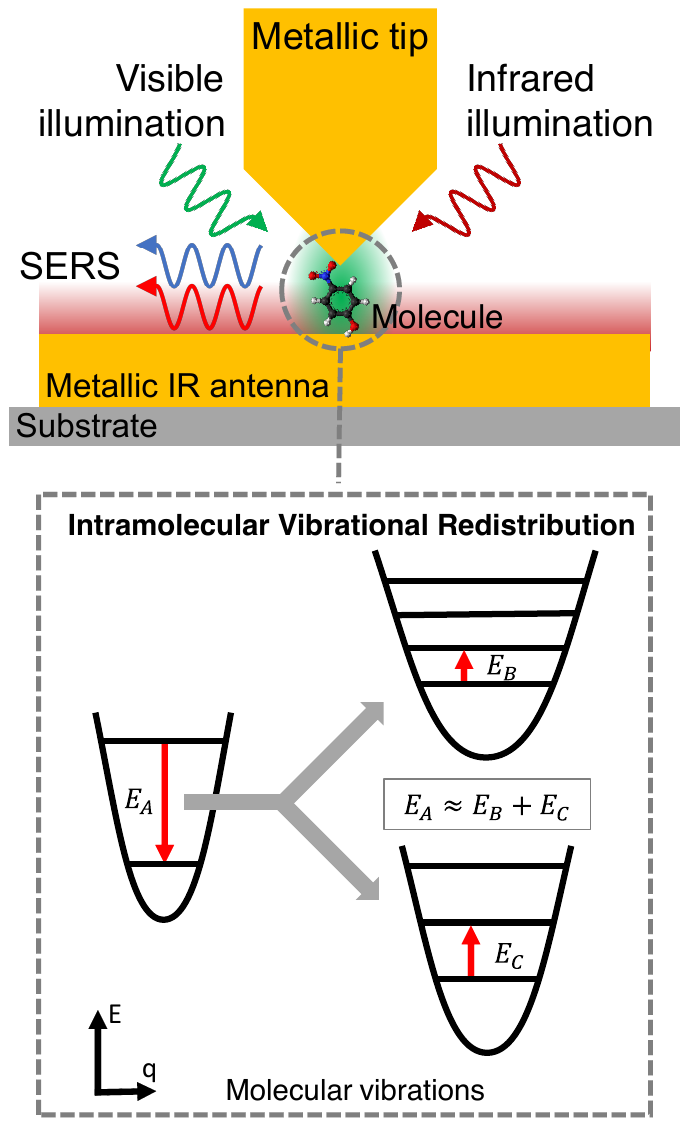}
    \caption{ \textit{Sketch of the system under study.}
    Top: Sketch of a single molecule inside a metallic nanoresonator. As a representative example of the system considered, a plasmonic nanocavity consisting in an ultrathin gap (typically $\sim$ 1 nm) between a metallic tip and a planar metallic antenna. The plasmonic gap nanocavity sustains modes in the visible range that can enhance the incident electric field (green shadowing in the figure). The IR antenna sustains a mode in the infrared range corresponding to molecular vibrations ($\sim$ 3-50 \textmu m), which can enhance an IR incident field (dark red shadowing in the figure). 
    Bottom inset: illustration of an intramolecular vibrational redistribution pathway inside the molecule, where an excited high energy vibration (left) relaxes by simultaneously transferring its energy to two other vibrations to which it is coupled (right).
     }
    \label{fig:schema}
\end{figure}

In this section, we introduce the theoretical framework to describe the coupling of a visible and an infrared (IR) cavity to a molecule, the illumination of this system by visible or infrared light and the vibrational energy redistribution within the molecule (see Fig. \ref{fig:schema}). We first describe the framework of molecular optomechanics adopted to model the Raman interaction between a cavity illuminated by light and molecular vibrations that do not couple with each other. We then review the origin of IVR and introduce the specific Fermi resonance pathway that we consider in the theoretical framework.

\subsection{Interaction between light, cavity modes and independent vibrations}\label{sec:description_system}

\subsubsection{Vibrational Hamiltonian}\label{sec:canonical_molecular_opto}

Throughout Section \ref{sec:description_system}, we consider a set of independent vibrations in a single molecule. Within a quantum mechanical description, the energy of molecular vibrations can be described with the Hamiltonian:
\begin{equation}\label{eq:free_vibrational_hamiltonian}
\hat{\mathcal{H}}_{\textsc{vib}}= \sum_i \hat{\mathcal{H}}_{\textsc{vib},i}  = \sum_i  \hbar \omega_{i} \hat{b}^{\dagger}_{i} \hat{b}_{i},
\end{equation}
where the sum runs over the $i$ vibrations of the molecule with frequency $\omega_{i}$, $\hbar$ stands for Planck's constant and $\hat{b}^{\dagger}_{i}$, $\hat{b}_{i}$ are respectively the Bosonic operators for creation and annihilation of vibrational excitations. We address next the coupling of this system to a cavity illuminated by light. 

\subsubsection{Visible illumination: molecular optomechanics description of SERS with independent vibrations}\label{sec:infrared_molecular_coupling}

We first model a standard configuration in SERS, as the one sketched in Fig. \ref{fig:schema}, where the vibrational modes of the single molecule interact inelastically with visible light through non-resonant Raman scattering. This interaction is strengthened by the presence of a locally large electromagnetic field (depicted by the green shadowing in the figure) induced by an optical nanoresonator, illustrated by a plasmonic gap nanocavity formed between a metallic tip and a flat metallic antenna (note that at visible wavelengths the flat metallic antenna behaves similarly to an infinite metallic substrate). To model this system, we use a recently developed quantum electrodynamical approach to SERS, referred to as molecular optomechanics due to its similarity with the framework of cavity optomechanics \autocite{roelli2016molecular,schmidt2016quantum,esteban2022molecular,roelli2024nanocavities}. In this molecular optomechanics approach, the total Hamiltonian is $\hat{\mathcal{H}}_{\text{tot}}(t)= \hat{\mathcal{H}}_{\textsc{vib}} + \hat{\mathcal{H}}_{\textsc{vis}}(t)$ where:
\begin{equation}\label{eq:hamiltonian_SERS_optomechanical}
\begin{split}
\hat{\mathcal{H}}_{\textsc{vis}}(t)= & \,\,\, \hbar \omega_{\textsc{vis}}^{\text{cav}} \hat{a}^{\dagger}_{\textsc{vis}} \hat{a}_{\textsc{vis}} \\
& + i\hbar \Omega_{\textsc{vis}}(t) ( \hat{a}^{\dagger}_{\textsc{vis}} e^{-i\omega_{\textsc{vis}} t} - \hat{a}^{}_{\textsc{vis}} e^{i\omega_{\textsc{vis}} t})  \\
& - \sum_i \hbar g_{\textsc{vis},i} \hat{a}^{\dagger}_{\textsc{vis}} \hat{a}_{\textsc{vis}} ( \hat{b}_i^{\dagger}  + \hat{b}_i^{} ).
\end{split}
\end{equation}

The first line on the right-hand side in Eq. (\ref{eq:hamiltonian_SERS_optomechanical}) describes the energy of the cavity, which is assumed to support a single-mode at visible frequency $\omega_{\text{vis}}^{\text{cav}}$, and $\hat{a}^{\dagger}_{\textsc{vis}}$ and $\hat{a}_{\textsc{vis}}$ are the Bosonic operators of creation and annihilation of the cavity photons. The second line in Eq. (\ref{eq:hamiltonian_SERS_optomechanical}) describes the time-dependent driving of the cavity by visible light with frequency $\omega_{\textsc{vis}}$ and with strength $\Omega_{\textsc{vis}}(t)\propto \sqrt{I_{\textsc{vis}}}(t)$ with $I_{\textsc{vis}}$ the illumination intensity (see Eq. (S2) of the SI). The third line stands for the optomechanical interaction  between the cavity and the vibrations, where the optomechanical coupling strength $g_{\textsc{vis},i}$ is inversely proportional to the cavity mode volume (see Eq. (S3) of the SI). The term $\propto  \hat{a}^{\dagger}_{\textsc{vis}} \hat{a}_{\textsc{vis}} \hat{b}_i^{\dagger}$ accounts for the Stokes Raman scattering process while the term $\propto  \hat{a}^{\dagger}_{\textsc{vis}} \hat{a}_{\textsc{vis}} \hat{b}_i$ accounts for the anti-Stokes Raman scattering process. Finally, we emphasize that in this standard molecular optomechanics approach the molecular vibrations are considered as independent. 

In molecular optomechanics, the time-evolution of the system described by Eq. (\ref{eq:hamiltonian_SERS_optomechanical}) can be calculated with a Lindblad master equation. In particular, the density matrix of the system $\hat{\rho}$ evolves as \autocite{schmidt2016quantum}:
\begin{equation}\label{eq:master_equation}
\begin{split}
    \partial_t \hat{\rho} = \frac{i}{\hbar} & \Big[\hat{\rho}(t), \hat{\mathcal{H}}_{\textsc{vib}} + \hat{\mathcal{H}}_{\textsc{vis}}(t) \Big] +  \frac{\kappa_{\textsc{vis}}}{2} \mathcal{D}_{\hat{a}_{\textsc{vis}}^{}}[\hat{\rho}] \\
    & + \sum_i \frac{\gamma_i (1 + n_i^{th})}{2} \mathcal{D}_{\hat{b}^{}_i}[\hat{\rho}] + \frac{\gamma_i n_i^{th}}{2} \mathcal{D}_{\hat{b}_i^{\dagger}}[\hat{\rho}],
\end{split}
\end{equation}
where we introduce the rate $\kappa_{\textsc{vis}}$ describing incoherent photon losses out of the cavity and the Lindblad‑Kossakowski terms $\mathcal{D}_{\hat{O}^{}}[\hat{\rho}] = 2\hat{O} \hat{\rho} \hat{O}^{\dagger} - \hat{O}^{\dagger} \hat{O} \hat{\rho} - \hat{\rho} \hat{O}^{\dagger} \hat{O}$. The rates $\gamma_i$ describe incoherent non-radiative population exchange between vibration $i$ and a phonon bath in thermodynamic equilibrium at temperature $T$. Hence, the thermal vibrational population is $n_i^{th}= 1/[\exp(\hbar \omega_i/k_B T) -1]$ with $k_B$ the Boltzmann constant. The rates $\gamma_i$ account phenomenologically for all the vibrational loss channels other than the IVR channels that we explicitly model in the rest of this work. In section S1.1 of the SI, we discuss several usual approximations used to solve Eq. (\ref{eq:master_equation}) with a low computational cost. In particular, under stationary illumination $\Omega_{\textsc{vis}}(t)=\Omega_{\textsc{vis}}^{\text{cw}}$, an analytical expression of the steady-state vibrational population $n_i^{\textsc{cw}}$ is available:
\begin{equation}\label{eq:vibrational_occupation_steady_state}
    n_i^{\text{cw}} = n_i^{th} + \frac{\Gamma_i^{+}}{\gamma_i},
\end{equation}
where $\Gamma_i^{+} \propto g_{\textsc{vis},i}^2 I_{\textsc{vis}}/[( \omega_{\textsc{vis}}- \omega_i - \omega_{\textsc{vis}}^{\text{cav}})^2 + (\kappa_{\textsc{vis}}/2)^2]$ is a rate of vibrational excitation by spontaneous Stokes SERS (see Eq. (S8) of the SI). Note that the expression given in Eq. (\ref{eq:vibrational_occupation_steady_state}) is valid for moderate pumping intensities when backaction from SERS remains negligible (see Section S1.1 of the SI). Notably, Eq. (\ref{eq:vibrational_occupation_steady_state}) shows that when $\Gamma_i^{+}/\gamma_i > n_i^{th} $, Stokes SERS drives the vibration out of thermal equilibrium. This nonequilibrium regime corresponds to the so-called vibrational pumping regime of SERS \autocite{maher2008vibrational,esteban2022molecular}, characterized by a linear scaling  of the population with pumping intensity. Finally, Eq. (\ref{eq:master_equation}) can also be solved to calculate Stokes and anti-Stokes SERS spectra (see section S1.1 of the SI). In particular, under continuous-wave pumping, the emitted intensity spectrally integrated over the anti-Stokes Raman line of a vibration $i$ can be written as:
\begin{equation}\label{eq:anti_stoke_SERS_semiclassical}
    I^{\text{aS}}_{i} \propto (\omega_{\textsc{vis}} + \omega_i)^4 \Gamma^{-}_i n_i^{\text{cw}},
\end{equation}
where $\Gamma^{-}_i\propto g_{\textsc{vis},i}^2 I_{\textsc{vis}}/[( \omega_{\textsc{vis}} + \omega_i - \omega_{\textsc{vis}}^{\text{cav}})^2 + (\kappa_{\textsc{vis}}/2)^2]$ is a rate of vibrational decay by spontaneous anti-Stokes SERS (see Eq. (S8) of the SI). This last equation shows that anti-Stokes SERS signals enable to get information on the evolution of vibrational populations $n_i^{\text{cw}}$ as a function of various parameters such as (slowly-varying) time or pumping.

In summary, visible illumination can be used to both populate a vibrational mode by Stokes SERS and to measure vibrational populations by anti-Stokes SERS. We will exploit this property throughout this work to study IVR.

\subsubsection{Infrared illumination of independent molecular vibrations}\label{sec:resonant_IR_pumping}

We consider next the coupling of a single molecule to a mid-infrared (hereafter infrared) resonator illuminated with infrared laser light.
The IR illumination pumps resonantly a given vibration $i$ of the same energy as the laser. In Fig. \ref{fig:schema}, the IR nanoresonator is illustrated by a metallic antenna (here a disk of a few \textmu m diameter) under the molecule \autocite{adato2015engineering}, which sustains IR plasmonic resonances that enhance the coupling between the laser and the vibrations. The elastic interaction of the vibration $i$ with the IR resonator is described by the Hamiltonian \autocite{roelli2020molecular}:

\begin{equation}\label{eq:hamiltonian_IR}
\begin{split}
    \hat{\mathcal{H}}_{\textsc{ir},i}(t)= & \hbar \omega_{\textsc{ir}}^{\text{cav}} \hat{a}_{\textsc{ir}}^{\dagger} \hat{a}_{\textsc{ir}} \\
    & + i \hbar \Omega_{\textsc{ir}}(t) ( \hat{a}_{\textsc{ir}}^{\dagger} e^{-i\omega_{\textsc{ir}} t} - \hat{a}_{\textsc{ir}}^{} e^{i\omega_{\textsc{ir}} t} ) \\
    & - \hbar g_{\textsc{ir},i} (\hat{a}_{\textsc{ir}}^{\dagger} \hat{b}_i +  \hat{a}_{\textsc{ir}} \hat{b}_i^{\dagger} ).
\end{split}
\end{equation}

The first line on the right hand-side in Eq. (\ref{eq:hamiltonian_IR}) describes the energy of the infrared cavity, which is assumed to support a single mode at frequency $\omega_{\textsc{ir}}^{\text{cav}}$, and has Bosonic creation and annihilation operators $\hat{a}_{\textsc{ir}}$ and $\hat{a}_{\textsc{ir}}^{\dagger}$. The second line in Eq. (\ref{eq:hamiltonian_IR}) describes the driving of the cavity by the IR light with frequency $\omega_{\textsc{ir}}$ and with strength $\Omega_{\textsc{ir}}(t)\propto \sqrt{I_{\textsc{ir}}}(t)$ with $I_{\textsc{ir}}$ the time-dependent IR illumination intensity. The third line stands for the interaction between the electric dipole of the vibration and the cavity electric field in the rotating wave approximation, with coupling strength $g_{\textsc{ir},i}$.

As in the previous section, the dynamics described by the Hamiltonian in Eq. (\ref{eq:hamiltonian_IR}) can be calculated in the framework of the Lindblad master equation (see Section S1.2 of the SI). In particular, under continuous-wave illumination $\Omega_{\textsc{ir}}(t)=\Omega_{\textsc{ir}}^{\text{cw}}$, the steady-state vibrational population can be written as:
\begin{equation}\label{eq:vibrational_occupation_steady_state_Ir_pumping}
    n_i^{\textsc{cw}} =  n_i^{th} + |\langle \hat{b}_i \rangle|^2,
\end{equation}
where $\langle \hat{b}_i \rangle \propto g_{\textsc{ir},i} \sqrt{I_{\textsc{ir}}}$ is the amplitude of the coherent vibrational population induced by the coherent laser pumping.

Equation (\ref{eq:vibrational_occupation_steady_state_Ir_pumping}) shows that when $|\langle \hat{b}_i \rangle|^2 \gtrsim n_i^{th}$, the vibration is driven out-of-equilibrium by the IR pumping. Similar to pumping by Stokes SERS, the induced population also scales linearly with pumping intensity in this situation. However, here IR pumping drives the vibration into a coherent state. In Section \ref{sec:IR_Raman_CW}, we exploit this pumping mechanism in a pump-and-probe configuration to study signatures of IVR.

\subsection{Intramolecular Vibrational Redistribution}\label{sec:IVR}

\subsubsection{Origin of IVR}

IVR refers to the ensemble of exchange processes between vibrational modes by which an excited molecule relaxes back to thermodynamic equilibrium. As sketched in the bottom inset of Fig. \ref{fig:schema}, a single microscopic IVR channel consists in coupling between two or several vibrations, such that population transfer between them becomes possible. To understand the origin and the form of these couplings, we focus on the Hamiltonian $\hat{\mathcal{H}}_{\text{mol}}$ describing the motion of $N$ independent vibrational degrees of freedom in the electronic ground state of a molecule \autocite{may2023charge}: $\hat{\mathcal{H}}_{\text{mol}} = \sum_{n=1}^N \hat{{P} }_n^2/2 + V(\boldsymbol{\hat{R}})$, where $\hat{{P}}_n$ is the (mass-weighted) momentum operator of the coordinate indexed by $n$ and $V(\boldsymbol{\hat{{R}}})$ is the potential energy surface of the molecule, which depends on the displacement operators $\boldsymbol{\hat{{R}}}=(\hat{R}_1,\dots, \hat{{R}}_N)$ of the coordinates from their equilibrium positions. Note that this expression of $\hat{\mathcal{H}}_{\text{mol}}$ relies on the Born-Oppenheimer approximation \autocite{may2023charge}. For weak absorption by a molecule with respect to the bond-dissociation energies, displacements of the coordinates around equilibrium are small so that a Taylor expansion of the potential energy surface can be made \autocite{may2023charge}, $V(\boldsymbol{\hat{{R}}}) = V^{(0)} + \sum_{n,m} V^{(2)}_{n,m } \hat{{R}}_n \hat{{R}}_m + \sum_{n,m,l} V^{(3)}_{n,m,l} \hat{{R}}_n \hat{{R}}_m \hat{{R}}_l + \dots$, where $V^{(M)}_{n_1,...,n_M} = [\partial^M V / \partial {R}_{n_1} ... \partial {R}_{n_M}](\boldsymbol{{R}}= \boldsymbol{0})/M! $ and $M!$ represents the factorial of M. The orthogonal basis in which the matrix $V^{(2)}_{n,m }$ is diagonal is called the normal mode basis. Injecting this Taylor expansion into $\hat{\mathcal{H}}_{\text{mol}}$ and expressing it in the normal mode basis yields \autocite{may2023charge}:\begin{equation}\label{eq:hamiltonian_molecule_normal_mode_basis}
    \hat{\mathcal{H}}_{\text{mol}} = \underbrace{\sum_{\phantom{,}n\phantom{l}}^{N} \frac{\hat{{p}}_n^2 }{2} +  \frac{1}{2} \omega_n^2 \hat{{q}}_n^2}_{\text{Independent harmonic modes}}  +  \underbrace{\,\,\sum_{n,m,l} U^{(3)}_{n,m,l} \hat{{q}}_n \hat{{q}}_m \hat{{q}}_l + \dots}_{\text{Anharmonic couplings}},
\end{equation}
where $\hat{{q}}_n$ is the normal mode coordinate of mode $n$, $\hat{{p}}_n$ its momentum and $\omega_n$ its frequency. The $U^{(M)}_{n_1,...,n_M}$ are linearly related to the $V^{(M)}_{n_1,...,n_M}$ by the transformation matrices defining the normal mode basis. The two first terms on the right-hand side of Eq. (\ref{eq:hamiltonian_molecule_normal_mode_basis}) correspond to the harmonic approximation, in which the vibrational modes do not couple to each other. These terms yield Eq. (\ref{eq:free_vibrational_hamiltonian}) when the creation and annihilation operators for the vibrational modes are introduced. However, the third term shows that in a realistic molecule the harmonic oscillators are coupled due to higher-order terms in the expansion of the potential energy surface. These anharmonic couplings are at the origin of IVR.

\begin{figure}[t]
    \centering
    \includegraphics[width= 0.50\textwidth]{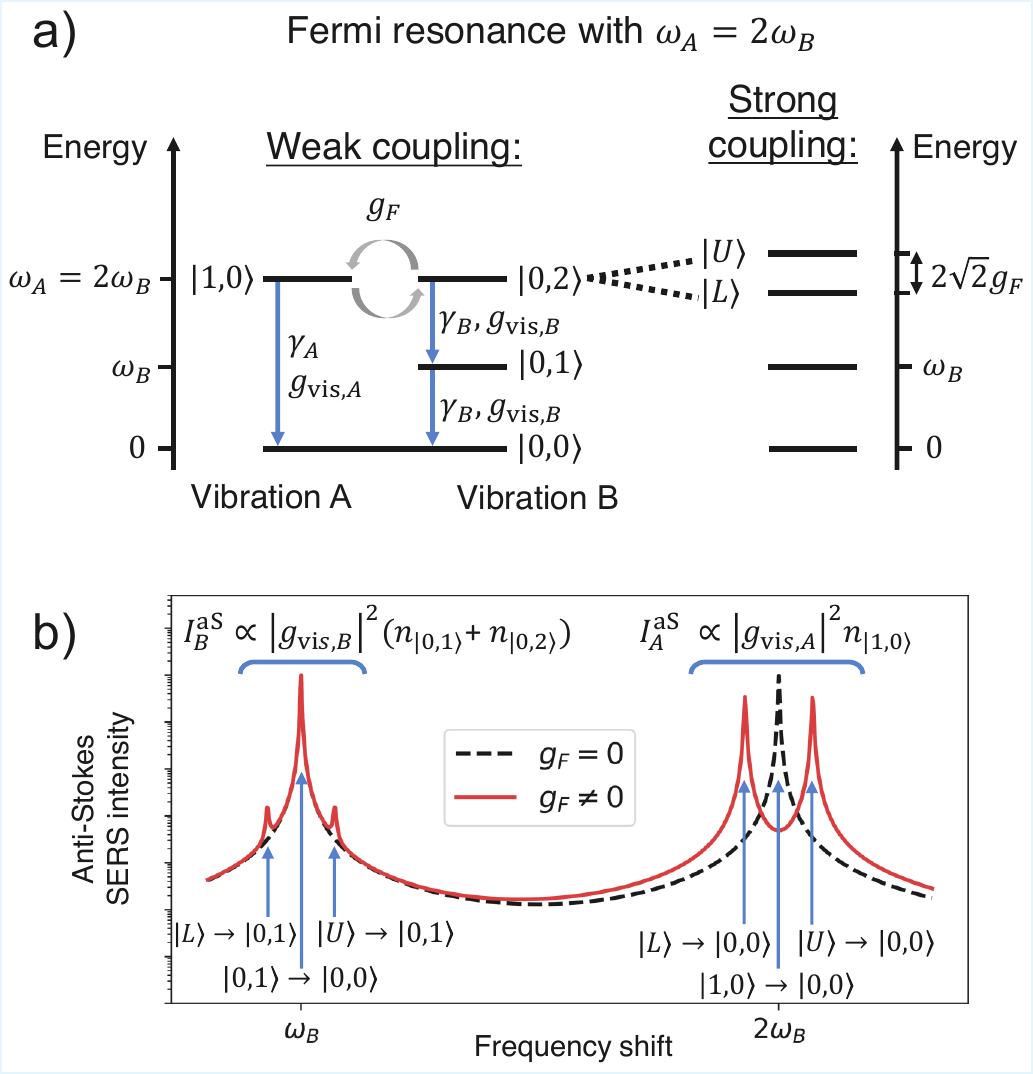}
    \caption{ \textit{Fermi resonance with $\omega_A=2\omega_B$.}
    (a): Schematic representation of the energy levels of two vibrations A and B in which the  first excited state of vibration A is resonantly coupled to the overtone of a vibration B by a Fermi resonance with coupling strength $g_{\textsc{f}}$ (gray arrows). The left part of the panel depicts the weak coupling point of view where modification of the energy levels by the coupling remains negligible ($2\sqrt{2}g_{\textsc{F}} \lesssim \gamma_A + \gamma_B$) while the right part depicts the strong coupling point of view where mode hybridization appears ($2\sqrt{2}g_{\textsc{F}} > \gamma_A + \gamma_B$). The blue arrows indicate decay channels, through non-radiative pathways at rates $\gamma_{\textsc{a}},\gamma_{\textsc{b}}$ or by anti-Stokes Raman scattering with optomechanical coupling strengths  $g_{\textsc{vis,a}},g_{\textsc{vis,b}}$. 
    (b): Sketch of an anti-Stokes SERS spectrum of two vibrations strongly coupled by a Fermi resonance (case $2\sqrt{2}g_{\textsc{F}} > \gamma_A + \gamma_B$, red solid line). The blue arrows indicate the possible Raman transitions. The equations on top indicate that the integrals of the anti-Stokes SERS intensity over the $\omega_B$ and $\omega_A$ peaks, respectively noted $I_{B}^{\text{aS}}$ and $I_{A}^{\text{aS}}$, are proportional to the product of the vibrational Raman coupling strengths square multiplied by vibrational populations. The black dashed line is the reference SERS spectrum for $g_{\textsc{F}}=0$. 
     }
    \label{fig:schema_Fermi_resonance}
\end{figure}

\subsubsection{Fermi resonance coupling}\label{sec:fermi_resonance_coupling}

The origin of Fermi resonances lies in the third-order anharmonic couplings $U^{(3)}_{n,m,l}$ in which the sum of the frequencies of two vibrations (or the second harmonic of one vibration) is close to resonance with the frequency of the third vibration (e.g. $\omega_n + \omega_m \approx \omega_l$) \autocite{fermi1931ramaneffekt,may2023charge}. Due to this resonance condition, Fermi resonances are very efficient relaxation pathways that yield a dominant contribution to IVR \autocite{karmakar2020intramolecular,poudel2024vibrational}. For the sake of simplicity, we focus hereafter on a specific Fermi resonance involving only two vibrational modes, labeled by A and B, in which the overtone of vibration B is resonantly coupled to the first excited state of vibration A, that is $U^{(3)}_{B,B,A}$ with $\omega_A \approx 2\omega_B$. The general case involving 3 vibrational modes $\omega_n + \omega_m \approx \omega_l$ is addressed in the SI. The configuration with 2 vibrations is sketched in Fig. \ref{fig:schema_Fermi_resonance} (a). Further, we will restrict this discussions to the first excited state of vibration A, noted $|1,0\rangle$, and the  first excited state $|0,1\rangle$ and overtone $|0,2\rangle$ state of vibration B, as we have checked numerically that the population of higher-order states are negligible for the pumping intensities considered in this work. Next, we expand in Eq. (\ref{eq:hamiltonian_molecule_normal_mode_basis}) the normal mode coordinate operators using the vibrational creation and annihilation operators, that is $\hat{{q}}_i = {q}_i^0 (\hat{b}^{\dagger}_i + \hat{b}_i)$ with ${q}_i^0= \sqrt{\hbar / 2 \omega_i}$ the zero-point amplitude of vibration $i$. 
The resulting Hamiltonian $\hat{\mathcal{H}}_{f}$ describing the coupling between vibrations through a Fermi resonance can be cast into the form $\hat{\mathcal{H}}_{\text{mol}}= \hat{\mathcal{H}}_{\textsc{vib}} + \hat{\mathcal{H}}_{\textsc{f}}$ with: 
\begin{equation}\label{eq:hamiltonian_Fermi_resonance}
\begin{split}
    \hat{\mathcal{H}}_{\textsc{f}} = & \hbar g_{\textsc{f}} (\hat{b}_A (\hat{b}_B^{\dagger})^2  + \hat{b}_A^{\dagger} \hat{b}_B^2 ),
\end{split}
\end{equation}
where $\hbar g_{\textsc{f}}= U^{(3)}_{B,B,A}  {q}_A^0 ({q}_B^0)^2$. The rotating wave approximation has been applied to obtain Eq. (\ref{eq:hamiltonian_Fermi_resonance}), so that fast rotating terms are discarded (e.g. $\hat{b}_A \hat{b}_B^2$ that  oscillates at frequency $\omega_A + 2 \omega_B$). The effect of the Hamiltonian in Eq. (\ref{eq:hamiltonian_Fermi_resonance}) is represented schematically in Fig. \ref{fig:schema_Fermi_resonance} (a). The left panel corresponds to weak coupling of the vibrations, which is defined by $2\sqrt{2}g_{\textsc{F}} < \gamma_A + \gamma_B$ (see next paragraph), and the gray arrows point at the fact that when one phonon is created on vibration A, two phonons are destroyed on vibration B, or the opposite. This exchange is most efficient when the resonance condition $\omega_A \approx 2\omega_B$ is satisfied. The right part of Fig. \ref{fig:schema_Fermi_resonance} (a) shows the strong coupling regime ($2\sqrt{2}g_{\textsc{F}} > \gamma_A + \gamma_B$), in which states $|1,0\rangle$ and  $|0,2\rangle$ hybridize into an upper (U) and a lower (L) mixed state,  respectively $|U\rangle = (|1,0\rangle + |0,2\rangle)/\sqrt{2}$ and $|L\rangle = (|1,0\rangle - |0,2\rangle)/\sqrt{2}$, that are separated by a frequency $2\sqrt{2} g_{\textsc{f}}$ (in the absence of losses, see also Section S1.3 of the SI).

Lastly, Fig. \ref{fig:schema_Fermi_resonance} (b) illustrates how state hybridization impacts the anti-Stokes Raman spectra in the strong coupling case. In particular, a schematic anti-Stokes SERS spectrum with strongly coupled vibrations ($g_{\textsc{f}} \neq 0$, red line) is compared to a spectrum with independent vibrations  showing the expected Raman peaks at frequencies $\omega_B$ and $\omega_A=2\omega_B$ ($g_{\textsc{f}}=0$, black dashed line). First, the peak associated to vibration A for $g_{\textsc{f}}=0$ splits into two peaks at frequencies $\sim 2\omega_B + \sqrt{2} g_{\textsc{f}}$ and $\sim 2\omega_B - \sqrt{2} g_{\textsc{f}}$ for $g_{\textsc{f}} \neq 0$. This is due to the component $\propto |1,0\rangle$ of the $|U\rangle$ and $|L\rangle$ hybrid states, that allows for $|U\rangle \to |0,0\rangle$ and $|L\rangle \to |0,0\rangle$ transitions induced by anti-Stokes Raman scattering processes. These two peaks in the Raman spectrum near $2\omega_B$ are referred to as a Fermi doublet \autocite{fermi1931ramaneffekt}. The doublet is expected to be observable provided that $2\sqrt{2}g_{\textsc{F}} > \gamma_A + \gamma_B$, which defines the strong coupling regime \autocite{torma2014strong}. Second, at frequencies near $\omega_B$, the component $\propto |0,2\rangle$ of the $|U\rangle$ and $|L\rangle$ states allows for transitions ($|U\rangle \to |0,1\rangle$ and $|L\rangle \to |0,1\rangle$) in strong coupling, while $|0,1\rangle \to |0,0\rangle$ transitions centered at $\omega_B$ are always possible and lead to the same contribution for $g_{\textsc{f}}=0$ and $g_{\textsc{f}}\neq 0$. As a consequence, 3 peaks with frequency near $\omega_B$ appear in the Raman spectrum in Fig. \ref{fig:schema_Fermi_resonance} (b) for $g_{_{F}}$ sufficiently large. 

Finally, the relation $I_{i}^{aS} \propto n_i$ derived for independent vibrations in Eq. (\ref{eq:anti_stoke_SERS_semiclassical}) remains valid within the weak coupling regime. The population of vibration A can be approximated by the population $n_{|1,0\rangle}$ of state $|1,0\rangle$, and the population of vibration B by the sum $n_{|0,2\rangle} + n_{|0,1\rangle}$ of the populations of state $|0,2\rangle$ and $|0,1\rangle$, respectively (note that we focus on the states population rather than the states occupation probability, so that $n_{|1,0\rangle} = \langle 1,0 | \hat{\rho} | 1,0\rangle$, $n_{|0,2\rangle} = 2 \langle 0,2 | \hat{\rho} | 0,2\rangle$ and $n_{|0,1\rangle} = \langle 0,1 | \hat{\rho} | 0,1\rangle$). Thus, $I^{aS}_A \propto g_{\text{vis},A}^2 n_{|1,0\rangle}$ and $I^{aS}_B \propto g_{\text{vis},B}^2 (n_{|0,1\rangle} + n_{|0,2\rangle})$, as also indicated in Fig. \ref{fig:schema_Fermi_resonance} (b). We use this scaling below to analyze the anti-Stokes SERS spectra.

\section{Results}\label{sec:results}

We next use the models established in Section \ref{sec:model} to study pump-and-probe measurements of IVR. We first introduce the model parameters used in our study. Then, we focus on a configuration where the vibrational pumping is induced by Stokes SERS processes, either under continuous-wave or pulsed illumination. We finally consider pumping by a cw IR laser. The main signatures of IVR observable by probing with spontaneous anti-Stokes SERS are highlighted for all these different pumping configurations.

\subsection{Parameters and approximations of the model}\label{sec:parameters}

To assess the feasibility of experimental demonstrations of the effects described below, it is important to set the study upon realistic plasmonic systems such as the one sketched in Fig. \ref{fig:schema}. In particular, we consider a metallic (typically silver or gold) tip separated by a $\sim$ 1 nm gap from a flat gold disk with a diameter in the micrometer range, as a configuration achievable in Scanning Probe Microscopy setups. We assume that a single molecule is placed in the gap. On the one hand, at visible frequencies, this arrangement sustains plasmonic resonances consisting of cavity modes characterized by extreme field confinement localized at the gap. For this reason, such plasmonic cavities are well-established platforms for efficient SERS \autocite{hoppener2024tip}. Note that the large disk behaves similarly as an infinite metallic substrate when analyzing the response at visible frequencies. On the other hand, at infrared (micrometer) wavelengths, the disk sustains plasmonic resonances capable of enhancing the IR electric field at its surface \autocite{adato2015engineering}, and thus the coupling of infrared light with the molecule. Finally, we assume that the single molecule and the tip are placed over a maximum of field of one of these infrared cavity modes, so that the molecule can be addressed by both visible and infrared light at the same time \autocite{chen2021continuous,xomalis2021detecting}, as will become relevant in Section \ref{sec:IR_Raman_CW}.

We introduce next realistic values for the parameters characterizing the system sketched in Fig. \ref{fig:schema}. We consider that the plasmonic cavity formed between the tip and the antenna has two modes in the visible range, labeled by 1 and 2, with respective frequencies $\hbar \omega_{\textsc{vis},1}^{\text{cav}}=$ 1.96 eV ($\sim$ 633 nm) and $\hbar \omega_{\textsc{vis},2}^{\text{cav}}=$ 1.58 eV ($\sim$ 785 nm) and equal losses for the two modes $\hbar \kappa_{\textsc{vis}}=$ 160 meV. These frequencies and linewidth are consistent with those obtained in experiments performed on plasmonic tips \autocite{zhang2017sub,roslawska2022mapping}. In addition, for both visible modes we consider a mode volume $\sim$ 300 nm$^3$ and an electric field enhancement of $\sim$ 250, that lie in the range of values that can be obtained in state-of-the-art plasmonic nanogap cavities \autocite{benz2016single,zhang2017sub,kongsuwan2020plasmonic,zhu2021influence,chen2021continuous}. With these values, the relation between the cavity illumination intensity and the cavity pumping strength in Eq. (\ref{eq:hamiltonian_SERS_optomechanical}) is $\hbar \Omega_{\textsc{vis}} [\text{meV}] = 1.13 \sqrt{ I_{\textsc{vis}} [\text{\textmu W.\textmu m}^{-2}] } $ for the visible cavity mode 1 and $\hbar \Omega_{\textsc{vis}} [\text{meV}] = 1.26 \sqrt{ I_{\textsc{vis}}[\text{\textmu W.\textmu m}^{-2}] } $ for the visible cavity mode 2 (see Eq. (S2) of the SI). 
We also consider that an infrared cavity mode due to the metallic disk is resonantly coupled to vibration A, so that $\hbar \omega_{\textsc{ir}}^{\text{cav}} = \hbar \omega_A$ (where the value of $\hbar \omega_A$ is given below). We consider the loss rate $\hbar \kappa_{\textsc{ir}}=$ 12 meV and the electric field enhancement $\sim$ 200 for this infrared cavity mode \autocite{xomalis2021detecting}. As a representative value for mid-infrared cavities \autocite{chen2021continuous}, we consider a mode volume of $\sim 10^4 \text{ nm}^3$, but we note that within our single-mode theory this quantity has no influence on the vibrational population $|\langle \hat{b}_i \rangle|^2$ induced by the IR pumping (see Eq. (\ref{eq:vibrational_occupation_steady_state_Ir_pumping}) and Eqs. (S18),(S19) of the SI). These values yield the scaling $\hbar \Omega_{\textsc{ir}} [\text{meV}] = 1.48 \sqrt{ I_{\textsc{ir}} [\text{\textmu W.\textmu m}^{-2}] }$ for the pumping of the infrared mode. 
We consider next that the molecule in the gap is a 4-nitrobenzenethiol (NBT) molecule (also called 4-nitrothiophenol), which is routinely studied in SERS experiments in plasmonic nanogap cavities \autocite{keller2018ultrafast,patil2023tip,bell2025coherent,xie2025continuous}. We specifically focus on the main $\omega_A \approx 2\omega_B$ Fermi resonance of this molecule, that involves its  carbon-sulfur stretching mode (mode A). Based on density functional theory (DFT) calculations of a single molecule attached to 19 atoms of gold (Au19NBT), we consider the vibrational frequencies $\hbar \omega_A = 2 \hbar \omega _B=$ 136 meV and a Fermi resonance coupling strength $\hbar g_{\textsc{f}}=$ 1.0 meV  (see discussion in Section S4.2 of the SI). To evaluate the Fermi resonance coupling strengths, we apply the deperturbed second-order vibrational theory approach \autocite{barone2005anharmonic} to the potential energy surface of Au19NBT.  We compute this potential energy surface within DFT using the exchange-correlation functional B3LYP-D3(BJ) \autocite{becke1993density,grimme2010consistent,grimme2011effect}, and the basis set 6-31G(d,p) \autocite{petersson1991complete} for the carbon, hydrogen, oxygen, nitrogen and sulfur atoms. In the case of the atoms of gold, we use the basis set LANL2DZ \autocite{hay1985ab}. We perform all the DFT calculations with the code Gaussian 16 Rev. B. 01 \autocite{g16}. We note that comparable values of $\hbar g_{\textsc{f}}$ have been found in similar molecules \autocite{poudel2024vibrational}. We also used DFT to calculate the Raman tensors and infrared dipoles of the selected vibrational modes in the Au19NBT configuration (see Section S4.1 of the SI for further discussion on the DFT calculations). By combining the obtained values for vibration A with the visible and IR cavity parameters (see Eqs. (S3) and (S19) of the SI), we get an optomechanical coupling strength of $\hbar g_{\textsc{vis,a}}=$ 0.021 meV for the cavity mode resonant at 633 nm (mode 1) and of $\hbar g_{\textsc{vis,a}}=$ 0.017 meV for the cavity mode resonant at 785 nm (mode 2), and the infrared coupling strength $\hbar g_{\textsc{ir,a}}=$ 0.018 meV. For vibration B, we obtain an optomechanical coupling strength $\hbar g_{\textsc{vis,b}}=$ 0.006 meV for mode 1 and $\hbar g_{\textsc{vis,b}}=$ 0.005 meV for mode 2. Additionally, we consider $\hbar\gamma_A = \hbar\gamma_B= $ 1.2 meV, which are decay rates consistent with the spectral width and the $<$ 2 ps relaxation times measured for several vibrational modes of NBT bound to gold \autocite{keller2018ultrafast}. Finally, we assume cryogenic or near-cryogenic temperature operation, at $T=$ 100 K in Sections \ref{sec:CW_IVR_SERS} and \ref{sec:pulsed_SERS}, and at $T=$ 150 K in Section \ref{sec:IR_Raman_CW} \autocite{meng2024local}. Indeed, to observe many of the effects reported in this work operating under ambient conditions, the use of extremely high-intensity powers would be required, which could destroy either the molecule, the metallic resonator, or both (see discussion in Section \ref{sec:discussion}).

Last, all the calculations of vibrational spectra and populations presented below are performed by considering usual approximations to the full master equation of the system (linearization of the optomechanical Hamiltonian and adiabatic elimination of the cavity degrees of freedom, see Sections S1.1 and S1.2 of the SI). These approximations yield a simplified master equation for the density matrix reduced to the vibrational modes. In Figs. \ref{fig:CW_SERS}, \ref{fig:pulsed_SERS} and \ref{fig:IR_SERS_Raman_upconversion}, this simplified master equation is solved numerically using the Python library QuTiP \autocite{johansson2012qutip,JOHANSSON20131234,lambert2024qutip} with 3 vibrational number states (i.e. states $|0\rangle$, $|1\rangle$, $|2\rangle$) for each vibration A and B. Within this level of theory, calculations with a higher number of vibrational number states (typically 10) are accessible and we checked that all the results shown in the figures are already converged with 3 number states. When possible, we have compared the results shown in Figs. \ref{fig:CW_SERS}, \ref{fig:pulsed_SERS} and \ref{fig:IR_SERS_Raman_upconversion} with results obtained by solving numerically the master equation without approximation. These comparisons are shown and discussed in detail in Section S2 of the SI. Overall, results from both implementations show very satisfactory agreement (the main difference is a larger background in calculated spectra when no approximation is made), supporting the validity of the approximations used. Finally, we emphasize that our model is general, so that it encompasses both the weak and strong coupling regimes of molecular vibrations coupled by a Fermi resonance (see further discussion in Section S1.3 of the SI).

\subsection{Continuous-wave Stokes SERS pumping}\label{sec:CW_IVR_SERS}

\begin{figure}[t]
    \centering
    \includegraphics[width= 0.7\textwidth]{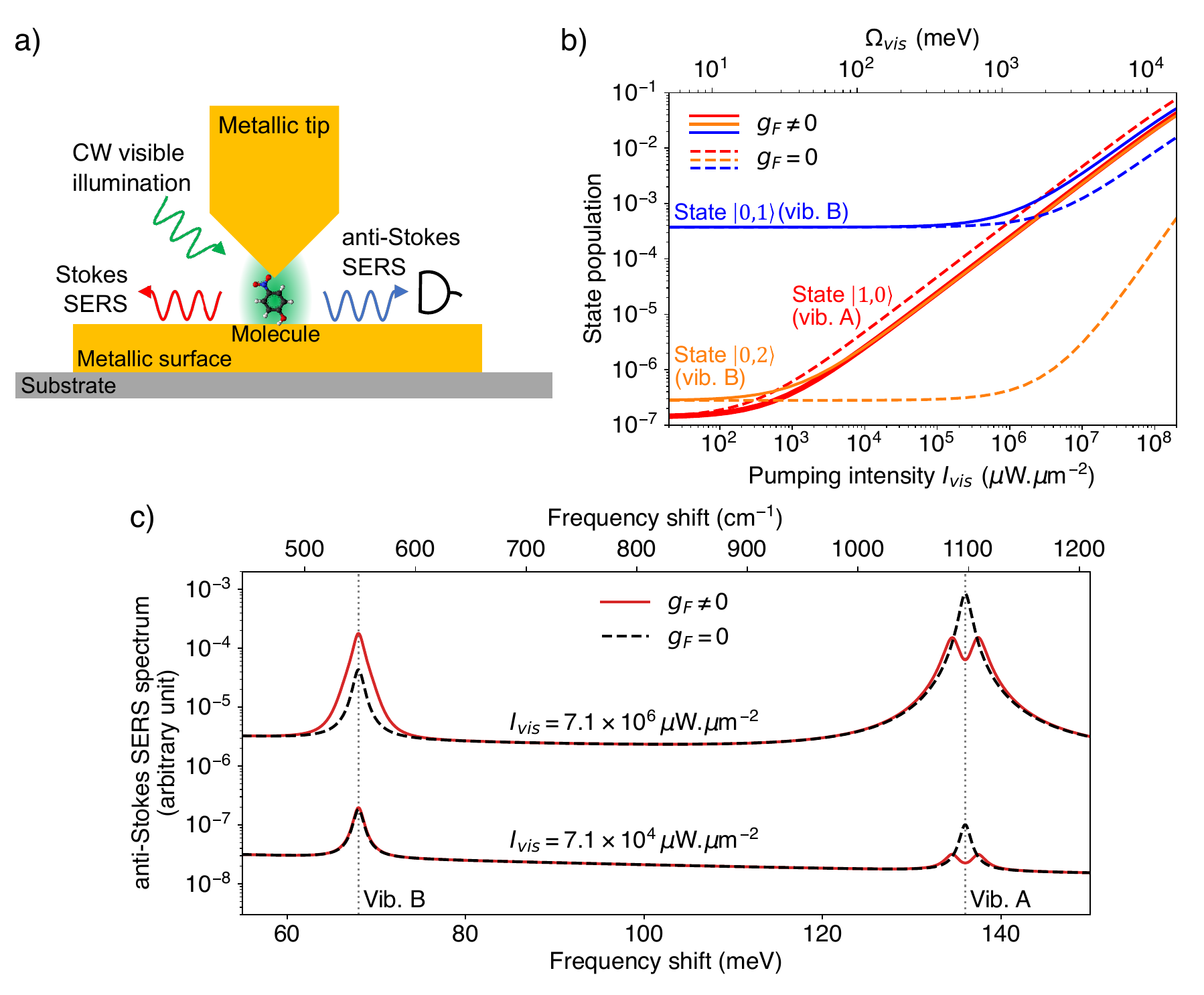}
    \caption{ \textit{Signatures of IVR under cw Stokes SERS pumping and cw anti-Stokes SERS probing}. 
    (a): sketch of the configuration under study, which consists in a single molecule in a plasmonic gap nanocavity with a mode resonant at 633 nm (mode 1 in Section \ref{sec:parameters}), resonantly illuminated by continuous-wave laser light (wavelength 633 nm, green arrow). The red arrow represents SERS emission by Stokes scattering processes, that also induces a pumping of the molecular vibrations. The blue arrow stands for the measured anti-Stokes SERS signal. We do not consider here IR illumination nor the IR cavity mode.
    (b):  Evolution of the population of state $|1,0\rangle$ (red continuous and dashed lines), state $|0,2\rangle$ (orange lines) and state $|0,1\rangle$ (blue lines) with the intensity of the pumping laser. The Fermi resonance coupling between states $|1,0\rangle$ and $|0,2\rangle$ is included ($g_F > 0$, solid lines) or neglected ($g_F =0$, dashed line).     
    (c): anti-Stokes SERS spectra of two molecular vibrations coupled by the Fermi resonance (red lines) or uncoupled (black dashed lines), for two pumping intensities $I_{\textsc{vis}}= 7.1\times10^4$ \textmu W.\textmu m$^{-2}$ ($\Omega_{\textsc{vis}}= $ 300 meV) and $I_{\textsc{vis}}= 7.1\times10^6$ \textmu W.\textmu m$^{-2}$ ($\Omega_{\textsc{vis}}=$ 3000 meV). The dotted vertical gray lines indicate the fundamental frequency of vibrations A and B. The temperature of the molecule is set to $T=$ 100 K in all panels. 
    The other parameters used in the simulations are introduced in Section \ref{sec:parameters}. 
    }  
    \label{fig:CW_SERS}
\end{figure}

We start by studying the pump-and-probe configuration for characterization of IVR sketched in Fig. \ref{fig:CW_SERS} (a), in which pumping is provided by a continuous wave (cw) visible laser inducing Stokes SERS and probing is provided by the same cw laser inducing anti-Stokes SERS. To this end, we consider the total Hamiltonian $\hat{\mathcal{H}}_{tot}(t) = \hat{\mathcal{H}}_{\textsc{vib,a}} + \hat{\mathcal{H}}_{\textsc{vib,b}} + \hat{\mathcal{H}}_{\textsc{f}} + \hat{\mathcal{H}}_{\textsc{vis}}(t)$ where $\hat{\mathcal{H}}_{\textsc{vib},i}$ is given in Eq. (\ref{eq:free_vibrational_hamiltonian}), $\hat{\mathcal{H}}_{\textsc{f}}$ is given in Eq. (\ref{eq:hamiltonian_Fermi_resonance}) and $\hat{\mathcal{H}}_{\textsc{vis}}$ is given in Eq. (\ref{eq:hamiltonian_SERS_optomechanical}). Here, $\hat{\mathcal{H}}_{\textsc{vis}}$ accounts for both pumping and probing with SERS under cw excitation at rate $\Omega_{\textsc{vis}}(t)=\Omega_{\textsc{vis}}^{\text{cw}}$. We assume that the visible light is resonant with the cavity mode 1 resonant at 633 nm, $\omega_{\textsc{vis}} = \omega_{\textsc{vis},1}^{\textsc{cav}}$, and neglect for simplicity the cavity mode 2 that is strongly non-resonant ($\omega_{\textsc{vis}} > \omega_{\textsc{vis},2}^{\textsc{cav}} + 2\kappa_{\textsc{vis}}$, see Section \ref{sec:parameters}). We further consider operation at cryogenic temperature $T=$ 100 K \autocite{meng2024local}. The procedure and approximations used to  obtain anti-Stokes SERS spectra and vibrational populations from the corresponding master equation in the steady-state are detailed in Section S2.1.1 of the SI, where the accuracy of the selected approach is compared to a rigorous approach as well as to a simplified semi-analytical model that reduces computational cost and provides additional insights. Calculated anti-Stokes SERS spectra are shown by the solid red lines in Fig. \ref{fig:CW_SERS} (c), where two pumping intensities $I_{\textsc{vis}}$ are considered. Data calculated with $g_{_F}=0$ are shown for reference (black dashed lines). We analyze below the spectral signatures caused by the Fermi resonance coupling between the vibrations.

For weak pumping $I_{\textsc{vis}}=7.1\times 10^{4}$ \textmu W.\textmu m$^{-2}$, the only effect of the Fermi resonance is to split the peak corresponding to vibration A (frequency shifts $\sim 2\omega_B$) into a doublet, as discussed in Section \ref{sec:fermi_resonance_coupling}. The doublet splitting is small in this case due to the moderate coupling strength of the Fermi resonance ($2\sqrt{2}g_{\textsc{F}} \approx \gamma_A + \gamma_B$). For the same reason, the three different contributions at frequencies around $\sim \omega_B$ that can be observed in Fig. \ref{fig:schema_Fermi_resonance} (b) merge into a single maximum in Fig. \ref{fig:CW_SERS} (c). This behavior illustrates the fact that the presence of a Fermi resonance can be difficult to identify only based on spectral fingerprints when $g_{\textsc{f}}$ is not sufficiently large, which is the most common case in realistic conditions. The presence of other vibrational modes at close frequencies can lead to accidental "doublet-like" lines that further complicate a spectral analysis. We also note that current theoretical tools for calculating the frequencies of vibrational modes have limited accuracy, which hinders the reliable identification of the origins of experimental Raman lines. Therefore, more reliable signatures of vibrational coupling are needed to gain insight into intramolecular vibrational redistribution. In particular, we analyze below how the SERS spectrum obtained under large pumping intensity $I_{\textsc{vis}} = 7.1\times 10^6$ \textmu W.\textmu m$^{-2}$ in Fig. \ref{fig:CW_SERS} (c) could serve to identify IVR.

To prepare the discussion of the emitted spectra under strong pumping and of the signatures of IVR that they contain, we analyze first the underlying physics of IVR under Stokes SERS pumping by focusing on the population of the states involved in the Fermi resonance. In Fig. \ref{fig:CW_SERS} (b), we show the evolution of the state population with pumping intensity $I_{\textsc{vis}}$. We first set the Fermi resonance coupling to zero ($g_{\textsc{f}}=0$), corresponding to the dashed lines. In this case, the population of the states $|0,1\rangle$ (blue dashed line) and $|0,2\rangle$ (orange dashed line) of vibration B is constant at low pumping intensity, corresponding to the thermal population. The population starts to increase with pumping for $I_{\textsc{vis}} \gtrsim 10^5$ \textmu W.\textmu m$^{-2}$. The population of the first excited state of vibration A, that is $|1,0\rangle$ (red dashed line), is also thermal at low intensity but starts to increase linearly with $I_{\textsc{vis}}$ at a lower threshold $I_{\textsc{vis}} \gtrsim 10^3$ \textmu W.\textmu m$^{-2}$. In the absence of IVR coupling, the increase in states population is due to the vibrational pumping induced by the optomechanical interaction between the visible pump, the optical cavity and the molecule, as discussed in Section \ref{sec:canonical_molecular_opto}. In particular, Eq. (\ref{eq:vibrational_occupation_steady_state}) shows that the population of the states becomes proportional to $g_{\textsc{vis},i}^2 I_{\textsc{vis}}$ when pumping induced by Stokes SERS processes overcomes non-radiative vibrational decay. The fact that $g_{\textsc{vis,a}} \gg g_{\textsc{vis,b}}$ (Section \ref{sec:parameters}) and the much lower thermal population of vibration A thus explains that the transition between thermal and nonthermal population is reached at much lower $I_{\textsc{vis}}$ for state $|1,0\rangle$ of vibration A than for states $|0,2\rangle$ and $|0,1\rangle$ of vibration B in the absence of Fermi resonance coupling.

We now introduce the Fermi resonance coupling  ($g_{\textsc{f}}>0$). In this case, both states $|1,0\rangle$ (red continuous line) and $|0,2\rangle$ (orange continuous line) start to scale linearly with $I_{\textsc{vis}}$ at $I_{\textsc{vis}} \approx 10^3$ \textmu W.\textmu m$^{-2}$, and the population of state $|1,0\rangle$ is moderately reduced (by $\approx$ 1/2) above this threshold as compared to the case $g_{\textsc{f}}=0$. These features point at a population transfer effect from state $|1,0\rangle$ to state $|0,2\rangle$ driven by the Fermi resonance coupling. In addition, the transition from thermal to nonthermal population of state $|0,1\rangle$ (blue continuous line) appears at lower pumping than in the case $g_{\textsc{f}}=0$, which points at a transfer of population from state $|0,2\rangle$ to state $|0,1\rangle$. We show in Section S2.1.1 of the SI that this latter transfer occurs due to non-radiative decay (i.e. through the term $\propto \gamma_B(1+n_B^{th})$ in Eq. (\ref{eq:master_equation})) and that $n_{|0,1\rangle}$ always remains $\geq n_{|0,2\rangle}$ for any value of the parameters considered. In summary, Fig. \ref{fig:CW_SERS} (b) shows that the Fermi resonance coupling enables a large population transfer from vibration A to vibration B. According to Eq. (\ref{eq:anti_stoke_SERS_semiclassical}) and given that $n_B= n_{|0,2\rangle} + n_{|0,1\rangle}$, an increase of population of vibration B leads to an enhancement of the anti-Stokes SERS emission from this vibration. Fig. \ref{fig:CW_SERS} (c) indeed shows that for $I_{\textsc{vis}} = 7.1\times10^6$ \textmu W.\textmu m$^{-2}$, the peak at frequency $\sim \omega_B$ (that corresponds to the anti-Stokes Raman scattering by vibration B) exhibits a large enhancement when $g_{\textsc{f}}>0$ (red line) as compared to $g_{\textsc{f}}=0$ (black dashed line). Hence, the population transfer induced by IVR can yield a signature on anti-Stokes SERS spectra.

In conclusion, the increase in the strength of anti-Stokes peaks induced by vibrational population transfer opens the possibility to characterize IVR pathways beyond the information contained in the difficult-to-interpret spectral positions of the Raman peaks. However, the identification of an IVR channel under continuous-wave pumping and probing still remains practically challenging. Indeed, in experiments, the coupled ($g_{\textsc{f}}>0$) and the uncoupled ($g_{\textsc{f}}=0$) configurations cannot be addressed separately, and thus compared. Further, the scaling with $I_{\textsc{vis}}$ of the population transferred through the Fermi resonance coupling shows the same linear dependency as the increase of population due to direct vibrational pumping, which prevents to use the scaling of the anti-Stokes SERS signal with pumping intensity to distinguish transfer by IVR from direct Stokes SERS pumping. In addition, continuous-wave illuminations as high as $I_{\textsc{vis}} =10^6$ \textmu W.\textmu m$^{-2}$ or more are unlikely to be achievable in experiments without destroying the system. Therefore, in the next section, we go beyond cw operation to investigate IVR signatures in time-resolved measurements under pulsed illumination.

\subsection{Pulsed Stokes SERS pumping}\label{sec:pulsed_SERS}

\begin{figure}[H]
    \centering
    \includegraphics[width= 0.7\textwidth]{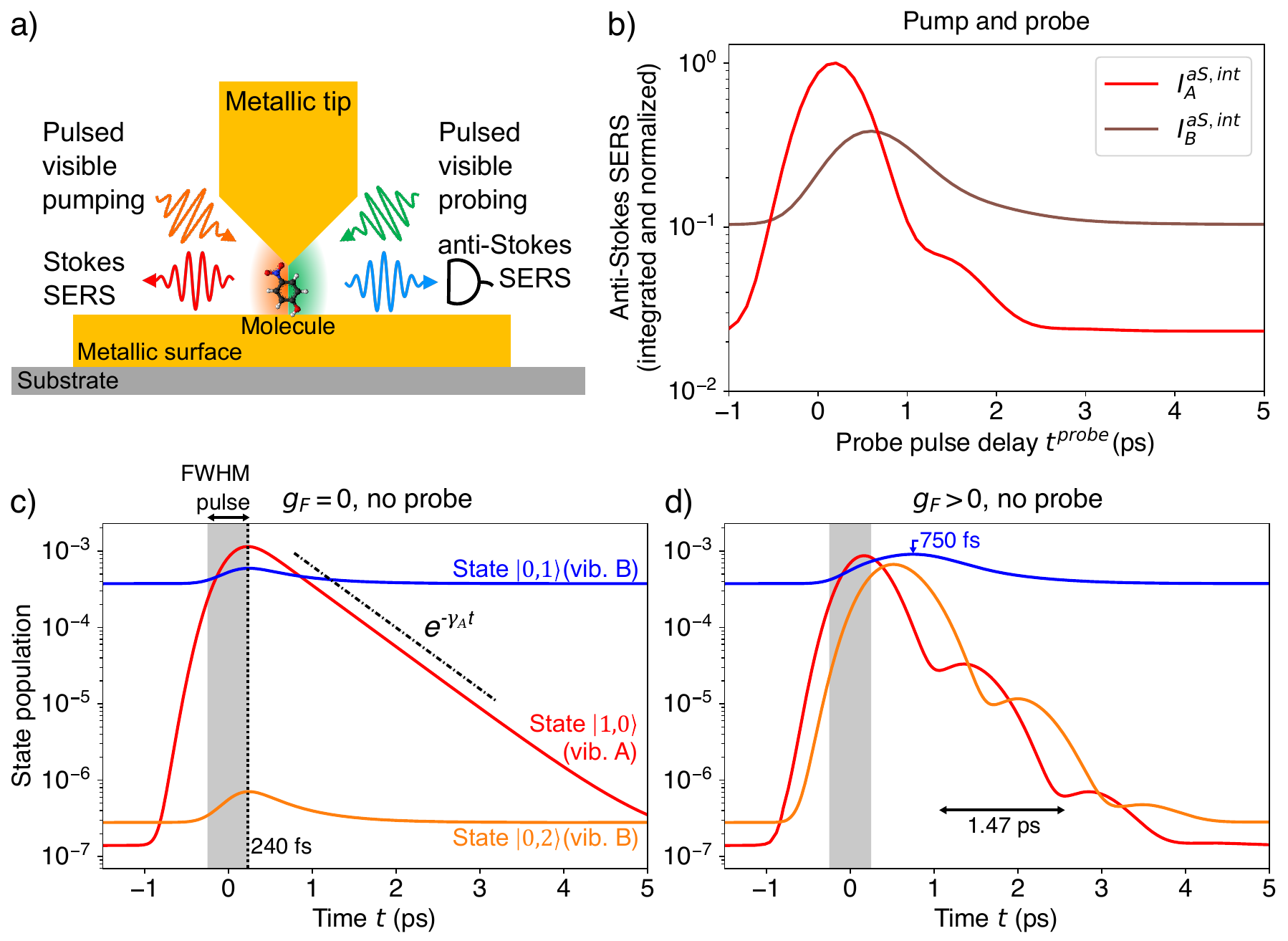}
    \caption{ \textit{Signatures of IVR under pulsed Stokes SERS pumping and pulsed anti-Stokes SERS probing}. 
    (a): Sketch of the configuration under study, which consists of a single molecule in a plasmonic gap nanocavity that supports two modes at visible wavelengths, one resonant at 633 nm (green shadowing on the figure, mode 1 in Section \ref{sec:parameters}) and another resonant at 785 nm (orange shadowing on the figure, mode 2 in Section \ref{sec:parameters}). The mode at 785 nm is resonantly illuminated with pulsed light (left orange arrow, pulse center at time $t=0$) that produces Stokes SERS light (red arrow) and vibrational pumping of the molecular vibrations. The cavity mode resonant at 633 nm is resonantly illuminated with pulsed light (right green arrow) arriving at a delayed time $t^{\text{probe}}$ that generates the measured anti-Stokes SERS signal (blue arrow). We do not consider here IR illumination nor the IR cavity mode.
    (b): Time dynamics of the frequency-integrated anti-Stokes SERS signals from molecular vibrations A and B coupled by a Fermi resonance $g_{\textsc{f}}>0$, $I^{\text{aS,int}}_{A}(t)$ (red line) and $I^{\text{aS,int}}_{B}(t)$ (brown line), respectively. Both curves are normalized to the maximum of $I^{\text{aS,int}}_{A}$.
    (c): Dynamics of the population of the states $|1,0\rangle$ (red line), $|0,2\rangle$ (orange line) and  $|0,1\rangle$ (blue line) without probe pulse ($\Omega_{\textsc{vis}}^{\text{probe,max}}=$ 0) and without Fermi resonance coupling ($g_{\textsc{f}}=0$). The vertical black dotted line indicates the time $\sim$ 240 fs at which all the states approximately reach their maximum population. The diagonal black dash-dotted line is a guide to the eyes showing an exponential relaxation of vibrational mode A with rate $\gamma_A$. The gray shaded area indicates the time interval during which the pump pulse intensity is greater than half its maximum value (FWHM). 
    (d): Same as in (c) but when the Fermi resonance coupling $g_{\textsc{f}}>0$ is included. 
    Parameters: in all panels the Gaussian pump pulse duration has a full-width at half maximum (of intensity) $\Delta\tau^{\text{pump}}=$ 500 fs and a peak intensity $I_{\textsc{vis}}^{\text{pump,max}}=$ 5.7$\times 10^{6}$\textmu W.\textmu m$^{-2}$ (i.e. $\Omega_{\textsc{vis}}^{\text{pump,max}}=$ 3000 meV). The Gaussian probe pulse is turned on in (b) with duration $\Delta \tau^{\text{probe}}=$ 500 fs and peak intensity $I_{\textsc{vis}}^{\text{probe,max}}=$ 1.3$\times 10^{5}$ \textmu W.\textmu m$^{-2}$ (i.e. $\Omega_{\textsc{vis}}^{\text{probe,max}}=$ 400 meV), and is turned off ($I_{\textsc{vis}}^{\text{probe,max}}=0$) in (c),(d). The temperature of the molecule is set to $T=$ 100 K. The other parameters used in the simulations are introduced in Section \ref{sec:parameters}.
    } 
    \label{fig:pulsed_SERS}
\end{figure}

We focus here on the signatures of IVR in time-resolved measurements with pulsed SERS pumping and probing (see Fig. \ref{fig:pulsed_SERS} (a)). To this end, we now consider the total Hamiltonian $\hat{\mathcal{H}}_{tot}(t, t^{\text{probe}} ) = \hat{\mathcal{H}}_{\textsc{vib,a}} + \hat{\mathcal{H}}_{\textsc{vib,b}} + \hat{\mathcal{H}}_{\textsc{f}} + \hat{\mathcal{H}}_{\textsc{vis}}^{\text{pump}}(t) + \hat{\mathcal{H}}_{\textsc{vis}}^{\text{probe}}(t-t^{\text{probe}})$, where the term $\hat{\mathcal{H}}_{\textsc{vis}}^{\text{pump}}(t)$ accounts for pumping induced by Stokes SERS processes with a pulse centered at time $t=0$ and $\hat{\mathcal{H}}_{\textsc{vis}}^{\text{probe}}(t-t^{\text{probe}})$ accounts for probing by anti-Stokes SERS with a pulse centered at a delayed time $t^{\text{probe}}$ (see Eq. (\ref{eq:hamiltonian_SERS_optomechanical}) for the expression of $\hat{\mathcal{H}}_{\textsc{vis}}$ and Eqs. (\ref{eq:free_vibrational_hamiltonian}) and (\ref{eq:hamiltonian_Fermi_resonance}) for the expressions of $\hat{\mathcal{H}}_{\textsc{vib},i}$ and $\hat{\mathcal{H}}_{\textsc{f}}$, respectively). 
Importantly, we consider that the pump pulse is tuned to the visible cavity mode resonant at 785 nm (mode 2 in Section \ref{sec:parameters}), i.e. $\omega_{\textsc{vis}} = \omega_{\textsc{vis},2}^{\textsc{cav}}$ in $\hat{\mathcal{H}}_{\textsc{vis}}^{\text{pump}}(t)$, and that the probe pulse is tuned to the cavity mode resonant at 633 nm (mode 1 in Section \ref{sec:parameters}), i.e. $\omega_{\textsc{vis}}^{} = \omega_{\textsc{vis},1}^{\textsc{cav}}$ in $\hat{\mathcal{H}}_{\textsc{vis}}^{\text{probe}}(t-t^{\text{probe}})$, and also remind that $\omega_{\textsc{vis},1}^{\textsc{cav}} > \omega_{\textsc{vis},2}^{\textsc{cav}} + 2\kappa_{\textsc{vis}}$. These choices guarantee that the SERS signals generated by the pump and probe pulses do not overlap spectrally and justify to neglect the driving of the cavity mode 1 by the pump beam and of the cavity mode 2 by the probe beam. These choices also ensure that the large Stokes photoluminescence by the metal constituting the plasmonic cavity does not overlap with the anti-Stokes SERS signal from the molecular vibrations induced by the probe pulse \autocite{cai2018photoluminescence,loirette2024theory}. 
We consider a Gaussian pump pulse in $\hat{\mathcal{H}}_{\textsc{vis}}^{\text{pump}}(t)$ given by $\Omega_{\textsc{vis}}^{\text{pump}}(t)= \Omega_{\textsc{vis}}^{\text{pump,max}} e^{-2\ln(2)\times[t/\Delta \tau^{\text{pump}}]^2}$ with $\Delta\tau^{\text{pump}}=$ 500 fs (full width at half-maximum of pulse intensity, FWHM) and $\Omega_{\textsc{vis}}^{\text{pump,max}}=$ 3000 meV (i.e. peak intensity $I_{\textsc{vis}}^{\text{pump,max}}= 5.7\times 10^{6}$ \textmu W.\textmu m$^{-2}$). The probe pulse is also Gaussian and follows $\Omega_{\textsc{vis}}^{\text{probe}}(t-t^{\text{probe}})= \Omega_{\textsc{vis}}^{\text{probe,max}} e^{-2\ln(2)\times[(t-t^{\text{probe}})/\Delta \tau^{\text{probe}}]^2}$ in $\hat{\mathcal{H}}_{\textsc{vis}}^{\text{probe}}(t-t^{\text{probe}})$ where $\tau^{\text{probe}}=$ 500 fs and $\Omega_{\textsc{vis}}^{\text{probe,max}}$ is either equal to 0 (no probing) or to 400 meV (i.e. $I_{\textsc{vis}}^{\text{probe,max}}= 1.3\times 10^{5}$ \textmu W.\textmu m$^{-2}$). As in the previous section, we consider operation at temperature $T=$ 100 K. The method used to obtain transient anti-Stokes SERS spectra and vibrational populations from the corresponding master equation is detailed in Section S2.2.1 of the SI. The resulting dynamics of anti-Stokes SERS signals and vibrational states populations are shown in Fig. \ref{fig:pulsed_SERS} and analyzed below.

We focus first on the transient dynamics of the vibrational populations under bare pulse pumping illumination (no probe, $\Omega_{\textsc{vis}}^{\text{probe,max}}=0$). Fig. \ref{fig:pulsed_SERS} (c) shows the transient dynamics of uncoupled vibrations ($g_{\textsc{f}}=0$). Initially, vibrational pumping induced by Stokes SERS processes leads to a peak population that is reached roughly at the same time $t\sim$ 240 fs for the three states, as indicated by the black dotted vertical line. Excitation of state $|1,0\rangle$ (i.e vibration A) is much stronger than that of states $|0,2\rangle$ and $|0,1\rangle$ (i.e vibration B), which is due to their different optomechanical coupling strength $g_{\textsc{vis,a}} \gg g_{\textsc{vis,b}}$. Once the pulse has declined, the states relax back to their thermal population, due to the non-radiative decay with rates $\gamma_A$ and $\gamma_B$, which results in an approximately exponential dependence of the population of state A with time ($\propto e^{-\gamma_A t}$ dependence marked by a dashed-dotted black line).

We analyze next in Fig. \ref{fig:pulsed_SERS} (d) the population transfer from vibration A to vibration B that occurs when $g_{\textsc{f}} > 0$. Interestingly, Fig. \ref{fig:pulsed_SERS} (d) shows coupled oscillations between the population of the  first excited state of vibration A ($|1,0\rangle$) and the overtone of vibration B ($|0,2\rangle$). This periodic population transfer can be interpreted as Rabi oscillations, similarly to the Rabi oscillations occurring between a two-level system and a cavity described in cavity quantum electrodynamics \autocite{grynberg2010introduction}. The period of the oscillations can be estimated by $\tau_F = \pi/\Omega_{\textsc{f}}=$ 1.47 ps (with $\Omega_{\textsc{f}}=\sqrt{2} g_{\textsc{f}}$ the Rabi splitting discussed in Section S1.3 of the SI), in good agreement with the numerical result shown in the figure. These Rabi-like oscillations mostly occur during the relaxation of the states toward their thermodynamical equilibrium, and they contrast strongly with the exponential decay observed for uncoupled vibrations in panel (c). Hence, these oscillations yield a clear signature of IVR. 

Another remarkable feature in Fig. \ref{fig:pulsed_SERS} (d) is the increase of population of the  first excited state of vibration B (state $|0,1\rangle$, blue line). This population increase is significantly larger than for $g_{\textsc{f}}=0$, and the maximum population is retarded with a certain delay, with a maximum at time $t\sim$ 750 fs against $t\sim$ 240 fs in panel (c). This increase and delay are induced by the Fermi resonance. In particular, the delay occurs because the population must be first transferred from the  first excited state of vibration A (i.e. state $|1,0\rangle$) to the overtone of vibration B (state $|0,2\rangle$) at rate $\sim \sqrt{2} g_{\textsc{f}}$, and then from $|0,2\rangle$ to $|0,1\rangle$ at rate $\gamma_B$. This sequence of transfers is much slower than the dynamics directly induced by optomechanical pumping related to instantaneous Stokes scattering processes. Hence, a delay of several hundreds of femtoseconds between the center of the pump pulse and the maximum population of a vibration yields another signature of the IVR pathway involving this vibration.

The changes in the vibrational populations can be monitored using the spontaneous anti-Stokes SERS signal generated by a weak probe pulse. To this end, we now consider a nonzero probe pulse tuned to the cavity mode  resonant at 633 nm (mode 1 in Section \ref{sec:parameters}) with $\Omega_{\textsc{vis}}^{\text{probe,max}}=$ 400 meV (i.e.  $I_{\textsc{vis}}^{\text{probe,max}}= 1.3\times 10^{5}$ \textmu W.\textmu m$^{-2}$) and pulse width $\Delta \tau^{\text{probe}}=$ 500 fs. A quantity accessible in experiments is the anti-Stokes SERS signal from the vibrational line $i$, that we calculate as $I_{i}^{\text{aS,int}}(t^{\text{probe}}) \propto  (\omega_{\textsc{vis,1}}^{\text{cav}} + \omega_i)^4 \int_{-\infty}^{+\infty} \Gamma_{1,i}^{-}[\Omega_{\textsc{vis}}^{\text{probe}}(t-t^{\text{probe}})] n_i(t) dt $ (see Section S2.2.1 of the SI), where $\Gamma_{1,i}^{-}$ is the rate of vibrational decay by anti-Stokes SERS in the cavity mode 1 (see Eq. (S55) of the SI). Fig. \ref{fig:pulsed_SERS} (b) shows the corresponding evolution of $I_{i}^{\text{aS,int}}(t^{\text{probe}})$ for vibration A (red line) and vibration B (brown line) as a function of the delay $t^{\text{probe}}$ between the probe and the pump pulse. Interestingly, the first period of the Rabi oscillations can be identified as a shoulder in the relaxation dynamics of $I_{A}^{\text{aS,int}}(t^{\text{probe}})$ (other periods cannot be observed because their signal falls below the SERS intensity produced by the probe pulse). Further, $I_{B}^{\text{aS,int}}(t^{\text{probe}})$ allows for identifying the long delay between the center of the pump pulse and the maximum population of the  first excited state of vibration B ($|0,1\rangle$). Hence, both signatures, Rabi oscillations and delay, could be accessible to experiments. Further discussion of the pumping intensity required to observe these signatures can be found in Section \ref{sec:discussion}. The influence of the probe pulse duration is discussed in Section S3 of the SI.

In summary, monitoring the transient dynamics of vibrational populations under pulsed Stokes SERS pumping and pulsed anti-Stokes SERS probing can provide evidence of signatures of IVR that might be accessible to experiments. The generality of these results is further highlighted in Section S2.2.2. of the SI, where we discuss the case $\hbar g_{\textsc{vis,a}}\sim \hbar g_{\textsc{vis,b}}$.

\subsection{Continuous-wave infrared pumping}\label{sec:IR_Raman_CW}

\begin{figure}[H]
    \centering
    \includegraphics[width= 0.7\textwidth]{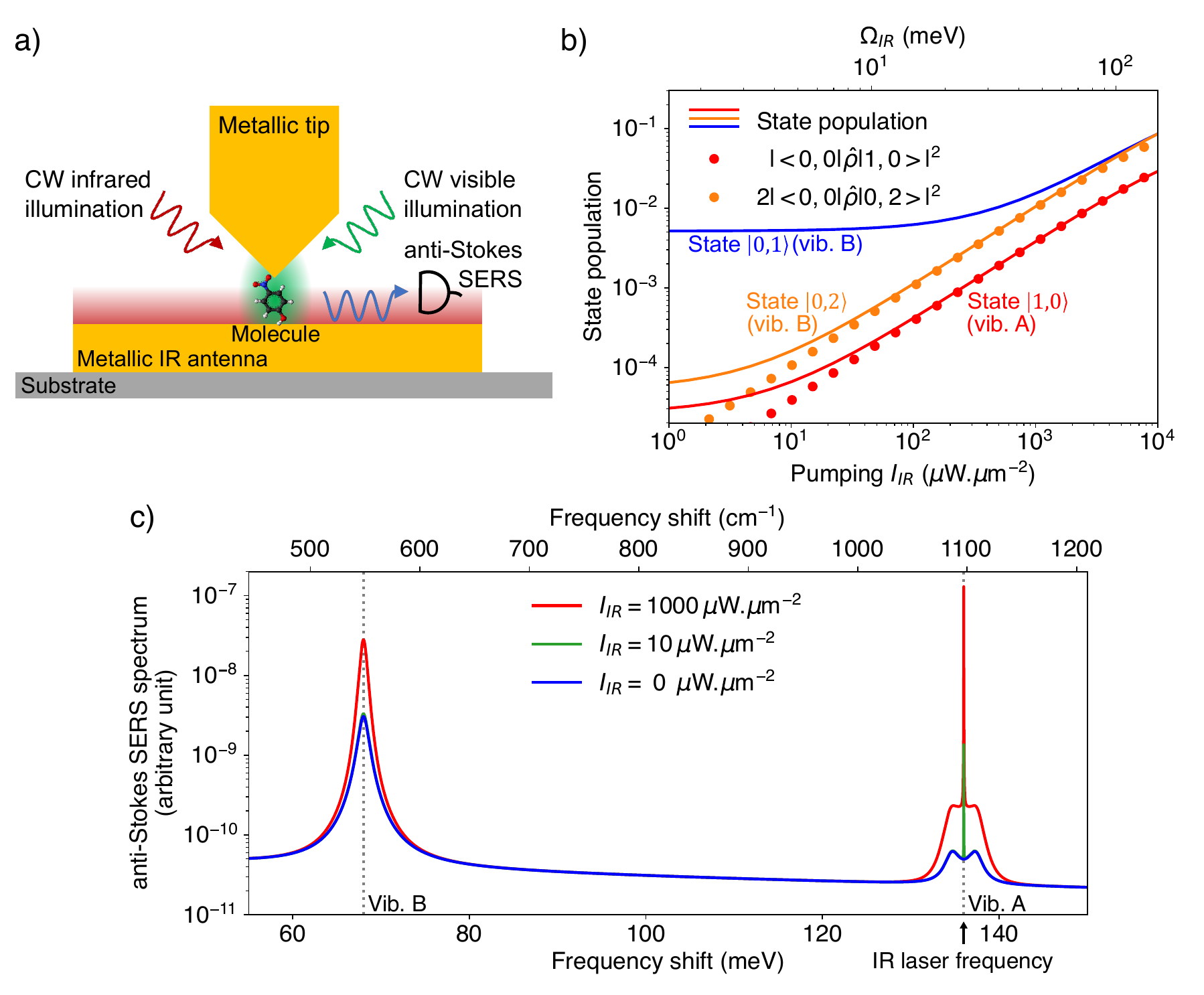}
    \caption{ \textit{Signatures of IVR under cw IR pumping and cw anti-Stokes SERS probing}. 
    (a): sketch of the configuration under study, which consists of a single molecule in a plasmonic gap nanocavity that supports a mode resonant at 633 nm (green shadowing on the figure, mode 1 in Section \ref{sec:parameters}). The bottom mirror of the nanocavity is a flat metallic disk that supports a plasmonic resonance in the infrared range (mode at 9.1 \textmu m, that is $\hbar \omega=$ 136 meV, IR mode in Section \ref{sec:parameters})  responsible for an enhancement of the IR electric field at the position of the molecule (red shadowing on the figure). The IR mode is resonantly illuminated with continuous-wave IR light (red arrow) that resonantly pumps a single molecular vibration of frequency $\hbar \omega_A=$ 136 meV. The visible cavity mode is resonantly illuminated by continuous-wave visible light (green arrow) that produces a measured anti-Stokes SERS signal revealing vibrational populations (blue arrow).
    (b): evolution of the populations of states $|1,0\rangle$ (red line), $|0,2\rangle$ (orange line) and $|0,1\rangle$ (blue lines) with the intensity of the pumping IR laser. The red dots show the quantity $|\langle 0,0|\hat{\rho}| 1,0 \rangle|^2$ that is the coherent population of state $|1,0\rangle$ (here it is almost equal to the coherent population $|\langle \hat{b}_A \rangle|^2$ of vibration A). The orange dots show the quantity $2|\langle 0,0|\hat{\rho}| 0,2 \rangle|^2$ that is the coherent population of state $|0,2\rangle$. 
    (c): anti-Stokes SERS spectra of the molecule for several cw IR pumping intensities $I_{\textsc{ir}}=$ 0 \textmu W.\textmu m$^{-2}$ (blue line), $I_{\textsc{ir}}=$ 10 \textmu W.\textmu m$^{-2}$ ($\Omega_{\textsc{ir}}=$ 4.7 meV, green line)) and $I_{\textsc{ir}}=$ 1000 \textmu W.\textmu m$^{-2}$ ($\Omega_{\textsc{ir}}=$ 46.8 meV, red line). The dotted vertical gray lines are guides to the eye and indicate the frequencies of vibration A and B.
    In panels (b) and (c), the Fermi resonance coupling is $g_{\textsc{f}}=$ 1.0 meV and the Raman probe intensity is $I_{\textsc{vis}}=$ 100 \textmu W.\textmu m$^{-2}$ ($\Omega_{\textsc{vis}}=$ 11.3 meV). The temperature of the molecule is set to $T=$ 150 K. The other parameters used in the simulations are introduced in Section \ref{sec:parameters}. In panel (c), the coherent contribution to the spectrum (nearly vertical red lines at 136 meV) is plotted using a Lorentzian lineshape with very small width (0.003 meV) and with an amplitude defined such that the spectral integral of the Lorentzian is equal to $(\omega_{\textsc{vis}} + \omega_A)^4 \Gamma^{-}_A |\langle 0,0|\hat{\rho}| 1,0 \rangle|^2$ (see Eq. (\ref{eq:anti_stoke_SERS_semiclassical})). The value 0.003 meV has been chosen to mimic the linewidth of the excitation laser.
    } 
    \label{fig:IR_SERS_Raman_upconversion}
\end{figure}

We finally explore an alternative pump-and-probe scheme in which vibration A is directly pumped by a cw infrared laser and the vibrations A and B are probed by cw anti-Stokes SERS (see Fig. \ref{fig:IR_SERS_Raman_upconversion} (a)). To this end, we consider the total Hamiltonian $\hat{\mathcal{H}}_{tot}(t) = \hat{\mathcal{H}}_{\textsc{vib,a}} + \hat{\mathcal{H}}_{\textsc{vib,b}} + \hat{\mathcal{H}}_{\textsc{f}} + \hat{\mathcal{H}}_{\textsc{ir,a}}^{}(t) + \hat{\mathcal{H}}_{\textsc{vis}}^{}(t)$ (see Eqs. (\ref{eq:free_vibrational_hamiltonian}),(\ref{eq:hamiltonian_SERS_optomechanical}),(\ref{eq:hamiltonian_IR}) and (\ref{eq:hamiltonian_Fermi_resonance})) in which $\omega_{\textsc{ir}} = \omega_{\textsc{ir}}^{\textsc{cav}}=\omega_A$ and $\Omega_{\textsc{ir}}(t) = \Omega_{\textsc{ir}}^{\text{cw}}$ in $\hat{\mathcal{H}}_{\textsc{ir,a}}(t)$, and $\omega_{\textsc{vis}} = \omega_{\textsc{vis},1}^{\textsc{cav}}$ (the other cavity mode resonant at frequency $\omega_{\textsc{vis},2}^{\textsc{cav}}$ can be neglected in this illumination and collection scheme) and $\Omega_{\textsc{vis}}(t) = \Omega_{\textsc{vis}}^{\text{cw}}=$ 11.3 meV ($I_{\textsc{vis}}=$ 100 \textmu W.\textmu m$^{-2}$ \autocite{xomalis2021detecting}) in $\hat{\mathcal{H}}_{\textsc{vis}}(t)$. We further consider operation at temperature $T=$ 150 K. The procedure and approximations used to obtain anti-Stokes SERS spectra and vibrational populations from the corresponding master equation are presented in Section S2.3.1 of the SI, together with additional semi-analytical insights to obtain the vibrational populations.  Calculated anti-Stokes SERS spectra are shown in Fig. \ref{fig:IR_SERS_Raman_upconversion} (c) for several IR laser pumping intensities $I_{\textsc{ir}}$. We analyze below the main characteristics of these spectra, focusing on the features that are connected with signatures of IVR.

The reference anti-Stokes spectrum in the absence of IR pumping is given by the blue line in Fig. \ref{fig:IR_SERS_Raman_upconversion} (c). This spectrum shows the same information as the one obtained in Fig. \ref{fig:CW_SERS} (c) for low pumping intensity. As already discussed in Section \ref{sec:CW_IVR_SERS}, it is challenging to identify an effect of IVR directly from this spectrum. We consider next an intermediate pumping intensity $I_{\textsc{ir}}=$ 10 \textmu W.\textmu m$^{-2}$ (green line). A first striking feature appears, which is a sharp and large peak at the frequency of the IR laser (136 meV). However, this peak is not related to any IVR process because it is also observed when vibrations are uncoupled ($g_{\textsc{f}}=0$). This peak appears because the IR laser pumping drives vibration A into a coherent state $\langle \hat{b}_A \rangle = \beta_A(\omega_{\textsc{ir}})$ and the visible laser pumping drives the optical cavity into a coherent state $\langle \hat{a}_{\textsc{vis}} \rangle = \alpha_{\textsc{vis}} (\omega_{\textsc{vis}})\propto \sqrt{I_{\textsc{vis}}}$ \autocite{kamandar2017quantum}. The optomechanical coupling Hamiltonian $g_{\textsc{vis},A} \hat{a}^{\dagger}_{\textsc{vis}} \hat{b}_{\textsc{a}} \hat{a}_{\textsc{vis}}$ in Eq. (\ref{eq:hamiltonian_SERS_optomechanical}) then turns into $g_{\textsc{vis},A} \hat{a}^{\dagger}_{\textsc{vis}} \beta_A(\omega_{\textsc{ir}}) \alpha_{\textsc{vis}} (\omega_{\textsc{vis}})$, which is formally the same Hamiltonian as the one considered in the configuration driven by a laser (see Eq. (\ref{eq:hamiltonian_SERS_optomechanical})). Accordingly, the term $g_{\textsc{vis},A} \hat{a}^{\dagger}_{\textsc{vis}} \alpha_{\textsc{vis}} (\omega_{\textsc{vis}}) \beta_A(\omega_{\textsc{ir}})$ drives the cavity field at the sum frequency $\omega_{\textsc{vis}}+\omega_{\textsc{ir}}$ into a coherent state, so that the resulting emission is coherent anti-Stokes SERS emission. This process corresponds to coherent vibrational sum-frequency generation \autocite{gruenke2016ultrafast}, and more information about this effect in the context of molecular optomechanics can be found in Refs. \autocite{roelli2020molecular,chen2021continuous,xomalis2021detecting,moradi2025photon,roelli2025operando}.

For large pumping intensities ($I_{\textsc{ir}}=$ 1000 \textmu W.\textmu m$^{-2}$, red line), in addition to the coherent peak, we observe a large enhancement of the broad (incoherent) peak at frequency shits $\sim \omega_B$ compared to the zero-pumping case. Figure \ref{fig:IR_SERS_Raman_upconversion} (b) shows that the origin of this incoherent peak is similar to the peak enhancement observed under cw SERS pumping (Section \ref{sec:CW_IVR_SERS}): first, the  first excited state of vibration A ($|1,0\rangle$) gets coherently populated by the direct IR laser pumping, as shown by the nearly perfect overlap between the total population (red line) and the coherent population $|\langle 0,0| \hat{\rho}|1,0\rangle |^2$ (red dots) of the state in the range $I_{\textsc{ir}}= 10^2 - 10^4$  \textmu W.\textmu m$^{-2}$. According to Eq. (\ref{eq:vibrational_occupation_steady_state_Ir_pumping}), the state population of $|1,0\rangle$ evolves linearly with  $I_{\textsc{ir}}$ in this range, as appreciated in the figure. A fraction of the population of state $|1,0\rangle$ is then transferred to state $|0,2\rangle$ due to the Fermi resonance coupling, inducing an out-of-equilibrium population of state $|0,2\rangle$ that also scales linearly with $I_{\textsc{ir}}$ (orange line and dots). As already evidenced by the Rabi oscillations found in Section \ref{sec:pulsed_SERS}, coherence is preserved during this transfer, which is indicated here by the almost perfect overlap between the population of state $|0,2\rangle$ (orange continuous line) and the coherent population $2|\langle 0,0 | \hat{\rho} | 0,2 \rangle|^2$ (orange dots) in the range $I_{\textsc{ir}}= 10^2 - 10^4$  \textmu W.\textmu m$^{-2}$. Finally, part of the population of $|0,2\rangle$ is transferred to $|0,1\rangle$ by non-radiative decay. We note that we obtain $|\langle 0,0 | \hat{\rho} | 0,1 \rangle = 0$ (not shown), so that coherence is lost during this last transfer. Ultimately, the increased population of state $| 0,1 \rangle$ leads to an increase of the anti-Stokes SERS signal at frequencies $\sim \omega_B$ because of $|0,1\rangle \to |0,0\rangle$ vibrational transitions, as shown in Fig. \ref{fig:IR_SERS_Raman_upconversion} (c). Given the incoherent population of state $| 0,1 \rangle$, the anti-Stokes SERS emission takes the form of a broad peak instead of a sharp line as obtained for coherent populations at frequency $2\omega_B$. Last, we note that the anti-Stokes SERS peak at frequencies $\sim \omega_B$ also receives a contribution from anti-Stokes Raman scattering processes involving $|0,2\rangle \to |0,1\rangle$ vibrational transitions. This contribution is also incoherent, because the Fermi resonance coupling only acts on the non-diagonal component $\langle 0,0 | \hat{\rho} | 0,2 \rangle$ of the density matrix but not on the component $\langle 0,1 | \hat{\rho} | 0,2 \rangle$ that is involved in the Raman scattering process, so that it adds to the broad anti-Stokes SERS peak at frequencies $\sim \omega_B$.

In summary, we have shown that a continuous-wave IR excitation resonant with the vibrational states composing the doublet of a Fermi resonance (near $\sim 2 \omega_B$) can lead to an excess of population of the low-energy states of the Fermi resonance (i.e. state $|0,1\rangle$) that can be detected in spontaneous anti-Stokes SERS. The signature of IVR found in this section is similar to the one obtained with cw Stokes SERS pumping, but here the possibility to selectively pump the Fermi doublet facilitates the assignment of the excess of signal to IVR. We also note that the use of pulsed IR illumination, not considered here, should also enable the characterization of the dynamics of the processes, similarly to that shown in Section \ref{sec:pulsed_SERS}. In addition, as discussed in Section \ref{sec:discussion}, the range of $I_{\textsc{ir}}$ used in our calculations can be implemented in realistic experiments. Therefore, the use of cw IR pumping in a dual-resonant (infrared and optical) resonator appears to be promising for the characterization of individual IVR pathways with identification of the vibrational modes involved.

\section{Discussion and conclusion}\label{sec:discussion}

In this work, we have evidenced signatures of IVR that can be observed with pump-and-probe vibrational spectroscopy. To obtain our results, we have resorted on several simplifications and hypotheses to focus on the main concepts. The paragraphs below aim to show that our conclusions are general and that the concepts developed in this work could be implemented within realistic  experimental platforms. A complementary discussion of other aspects (influence of probe pulse duration, the selectivity of vibrational mode pumping by Stokes SERS, the effect of competing signals on SERS measurements, the influence of collective effects due to the presence of multiple molecules, IVR of low energy vibrations and further considerations on molecular anharmonicities) can be found in Section S3 of the SI. 

First, we stress that Fermi resonances yield the dominant contribution to IVR and more generally to vibrational energy relaxation. For example, it has been  shown recently that accounting for all the main Fermi resonances of a molecule enables the recovery of experimentally measured vibrational lifetimes \autocite{poudel2024vibrational}. In this context, we have considered the simplest case of two vibrations A and B with perfect tuning $\omega_A = 2\omega_B$. For completeness, the general case of three vibrations A, B, C with imperfect tuning $\omega_A \approx \omega_B +\omega_C$ is studied in Sections S2.1.2, S2.2.3, S2.3.2 of the SI, showing that the same signatures of IVR are recovered. In addition, we note that our theoretical treatment can be readily extended to account for higher-order anharmonic couplings, such as Darling-Dennison resonances \autocite{darling1940water}, that could play an important role in some specific cases \autocite{zdanovskaia2022eight}.

Further, it is worthwhile to discuss the pumping strengths that we consider and the possibility to detect the emitted signals. For pulsed Stokes SERS pumping, we considered in Section \ref{sec:pulsed_SERS} pulses with 500 fs duration (FWHM) and peak intensity $I_{\textsc{vis}}^{\text{pump,max}} = 5.7\times10^6\,$\textmu W.\textmu m$^{-2}$, which corresponds to a fluence per pulse of 3 pJ.\textmu m$^{-2}$. Such fluence values have been used in SERS experiments in ultrathin gap nanocavities without introducing significant sample degradation \autocite{pozzi2016operational,jakob2023giant}. On the other hand,  a rough estimate  of the  emitted power (Section S3.3.A of the SI) indicates that experimental measurements of the signatures of IVR under pulsed visible probing on single molecules is challenging and might require additional optimization (see below), or alternatively, to perform the study on a few molecules ($\sim$ 1000), which would still provide a crucial advantage as compared to the large number of molecules required to identify IVR in more standard vibrational spectroscopy techniques \autocite{rubtsov2019ballistic}.  Therefore, this pumping configuration, combined with cryogenic temperature operation $T\lesssim$ 100 K, could enable the observation of signatures of IVR within realistic  experimental platforms. For continuous-wave IR illumination, Fig. \ref{fig:IR_SERS_Raman_upconversion}  (c) shows that vibrational redistribution effects should be clearly observable for $I_{\textsc{ir}} \approx$ 1000 \textmu W.\textmu m$^{-2}$, and Fig. \ref{fig:IR_SERS_Raman_upconversion}  (b) suggests that these effects could be already seen for $I_{\textsc{ir}}\gtrsim$ 100 \textmu W.\textmu m$^{-2}$. Pumping strengths up to $I_{\textsc{ir}}\approx$ 3000 \textmu W.\textmu m$^{-2}$ have recently been reported in reference \autocite{xie2025continuous}, which is based on an experimental platform comparable to the one studied in this work. In addition, an estimate of the anti-Stokes SERS intensity emitted under cw visible probing with intensity $I_{\textsc{vis}}\approx$ 100 \textmu W.\textmu m$^{-2}$ supports the possibility of detecting IVR in a single molecule under this illumination scheme (see Section S3.3.B of the SI). Therefore, this pumping and probing configuration, combined with temperatures typically between $T\approx$ 100 K and $T\approx$ 200 K (see also Section S2.3.2 of the SI), can also allow for observing signatures of IVR within current experimental platforms down to the single-molecule level.

Finally, we emphasize that the experimental configuration that we have studied could be optimized in several ways. For example, pumping induced by Stokes SERS processes should be more efficient in plasmonic cavities formed between a surface and an atomic protusion, the so-called picocavities \autocite{benz2016single,baumberg2022picocavities}, which can shrink the effective mode volume and thus increase the optomechanical coupling strength. The optomechanical coupling strength may also be improved using atomic antennas as proposed recently \autocite{schmidt2025molecular}. Additionally, resonant SERS excitation of molecular species that feature an electronic transition in the visible range could provide an alternative and efficient vibrational pumping mechanism \autocite{neuman2019quantum}. Last, more sophisticated infrared cavity designs could improve the efficiency of infrared excitation \autocite{chikkaraddy2023single,roelli2025operando}.

In conclusion, we have presented a theoretical framework based on cavity quantum electrodynamics that addresses the characterization of intramolecular vibrational redistribution within surface-enhanced pump-and-probe vibrational spectroscopy. A key result of our work is the generalization of the molecular optomechanics description of SERS to include the effect of anharmonic couplings between the vibrations of a molecule that gives rise to IVR. In the future, this framework could be extended to accurately model more sophisticated pump and probe schemes involving, e.g., pulsed sum-frequency generation \autocite{bell2025coherent} or stimulated Raman scattering \autocite{jakob2024accelerated}. Another important result of our work is the application of our framework to two relevant pump-and-probe configurations, based on infrared and visible pumping, respectively, and on probing with incoherent anti-Stokes SERS. For both configurations, we have identified clear signatures  that enable the characterization of individual IVR pathways. Our results suggest that these signatures could be accessible in a single molecule under  realistic experimental conditions. We thus believe that this work can foster and guide the development of a new generation of vibrational spectroscopy experiments able to shed light on the complex relaxation dynamics of vibrations in molecules.

\section*{Acknowledgements}

R.A.B. is thankful for the technical and human support provided by the Donostia International Physics Center (DIPC) Computer Center.

\subsection*{Funding Sources}

We acknowledge support from grant no. PID2022-139579NB-I00 funded by ERDF/EU and by MICIU/AEI/10.13039/501100011033 and from grant no. IT 1526-22 funded by the Department of Science, Universities and Innovation of the Basque Government. J.A. acknowledges financial support from the Elkartek project u4smart funded by the Department of Industry, Energy Transition and Sustainability of the Basque Government.  J.A. and R.A.B acknowledge financial support from the Elkartek project smartus funded by the Department of Industry, Energy Transition, and Sustainability of the Basque Government. A.L.-P acknowledges support from the Juan de la Cierva fellowship no. JDC2024-054665-I funded by MCIU/AEI/10.13039/501100011033 and by the ESF+. R.A.B. acknowledges the financial support from the grant awarded to the QSEIRA project (2024-QUAN-000011-01) from the Gipuzkoa Quantum program's 2024 call of the Department of Economic Promotion and Strategic Projects of the Provincial Council of Gipuzkoa.

\section*{Associated content}

\subsection*{Supporting information}

Section S1: Details of the derivation of the theoretical framework. Section S2: Further details on the modeling of the different pump-probe configurations for IVR
characterization studied in Section \ref{sec:results}. Section S3: Complementary discussion to Section \ref{sec:discussion} about the experimental feasibility of IVR detection using SERS platforms. Section S4: Details of the DFT calculations of the vibrational properties.

\subsection*{Data Availability Statement}

The data that support the findings of this article are openly available at DOI: \url{https://digital.csic.es/handle/10261/429235}.

\printbibliography

\end{document}